\documentclass[10pt, doublecolumn]{IEEEtran}
\usepackage{float}
\usepackage{array}    
\usepackage{tikz}     
\usepackage{amsmath}  
\usepackage{makecell} 
\usepackage{diagbox}  
\usepackage{dsfont}
\usepackage{graphicx}
\usepackage{stfloats}
\usepackage{caption}
\usepackage[ruled]{algorithm2e}
\ifCLASSOPTIONcompsoc
  \usepackage[caption=false,font=normalsize,labelfont=sf,textfont=sf]{subfig}
\else
  \usepackage[caption=false,font=footnotesize]{subfig}
\fi
\usepackage{indentfirst}
\usepackage{graphicx}
\usepackage{bm}
\usepackage{amsmath}
\allowdisplaybreaks[4]
\usepackage{amssymb}
\usepackage{times}
\usepackage{mathtools}
\usepackage{psfrag}
\usepackage{cite}
\usepackage{lastpage}
\usepackage{fancyhdr}
\usepackage{color}
 \usepackage{amsthm}
\usepackage{bigints}
\newtheorem{Theorem}{Theorem}
\newtheorem{Lemma}{Lemma}
\theoremstyle{remark}
\newtheorem{Remark}{$\quad$Remark}
\newtheorem{Corollary}{Corollary}

\begin{document}
\title{Age of Information Analysis for NOMA-Assisted Grant-Free Transmissions with Randomly Arrived Packets}
\author{Yanshi Sun, \IEEEmembership{Member, IEEE}, Yanglin Ye, Caihong Kai, Zhiguo Ding, \IEEEmembership{Fellow, IEEE}, Bin Chen \IEEEmembership{Senior Member, IEEE}
\thanks{
The work of Y. Sun and Y. Ye was supported in part by the National Natural Science Foundation of China under Grant 62301208.

Y. Sun, Y. Ye, C. Kai and B. Chen are  with the School of Computer Science and Information
Engineering, Hefei University of Technology, Hefei, 230009, China. (email: sys@hfut.edu.cn, yanglinyeuna@163.com, chkai@hfut.edu.cn, bin.chen@hfut.edu.cn).

Z. Ding is with the Department of Electrical and Electronic Engineering,
the University of Manchester, Manchester, Manchester M13 9PL, U.K. (e-mail:
zhiguo.ding@manchester.ac.uk).
}\vspace{-2em}}
\maketitle
\begin{abstract}
This paper investigates the application of non-orthogonal multiple access (NOMA) to grant-free transmissions to reduce the age of information (AoI) in uplink status update systems, where multiple sources upload their {status updates} to {a common} receiver. 
Unlike existing studies which {adopted} the idealized generate-at-will (GAW) model, {i.e., a status} update data can be generated and transmitted at any time, this paper utilizes a more practical model {to characterize} the inherent randomness
of the generation of the status updating data packets.
A rigorous analytical framework is established to precisely evaluate the average AoI achieved by the NOMA-assisted grant-free schemes for both {the} cases with and without retransmission. The impact of the choice of the probability {of transmission} on the average AoI is investigated.
Extensive simulation results are provided to validate the accuracy of the developed analysis. 
It is shown that NOMA-assisted schemes are more superior in reducing AoI{, compared} to orthogonal multiple access (OMA) based schemes. 
In addition, compared to schemes without retransmission, the AoI performance {of} the schemes with retransmission can {be improved} significantly when the status update generation rate is low or the user density is relatively high. 
\end{abstract}
\begin{IEEEkeywords}
Non-orthogonal multiple access (NOMA), age of information (AoI), grant-free, timeliness, retransmission, status updates, Markov chain
\end{IEEEkeywords}

\section{Introduction}
In many real-time applications, such as autonomous driving, industrial IoT, and remote healthcare, the freshness of information is critical to ensuring the reliability and performance of {these applications}. For example, in autonomous driving systems, outdated sensor data can lead to incorrect decisions, potentially resulting in accidents. Although traditional communication metrics, such as {system throughput} and end-to-end delay, have been widely used to evaluate network performance, they {are inadequate to capture} the timeliness or freshness of information. To address this gap, the concept of age of information (AoI) was introduced as a novel metric to quantify information freshness from the perspective of a receiver  \cite{kaul2012real, yates2021age,abd2019role}. AoI measures the time elapsed since the generation of the latest {successfully received} update, providing a more comprehensive understanding of the timeliness of information in real-time systems \cite{sun2017update,abbas2021markovian,costa2016age,Angelakis2017}.

The impact of multiple access (MA) techniques on AoI has raised significant attention, due to {the challenge that the scarce wireless spectrum needs} to be shared among a tremendous number of {users} in the network \cite{yates2018age,tse2005fundamentals}. Generally, MA techniques can be classified into two categories, namely static MA techniques and random MA techniques, respectively\cite{goldsmith2005wireless}. In static MA techniques, resource allocations for the {users} are pre-defined. In \cite{pan2020information,moltafet2020age,han2021fairness}, performance analysis and resource allocation optimization for the AoI achieved by
transmission schemes based on orthogonal multiple access (OMA) techniques {were} studied. It is further shown by \cite{ding2023age, sun2024age,liu2021peak} that by applying non-orthogonal multiple access (NOMA) techniques \cite{ding2017application}, the average AoI can be significantly reduced{, compared} to OMA based schemes. 
In contrast to static MA based schemes, random MA based schemes, such as ALOHA and carrier sensing MA (CSMA) \cite{goldsmith2005wireless}, are better suited for scenarios with dense nodes which have sporadic traffics. 
AoIs achieved by slotted ALOHA and CSMA networks were analyzed in \cite{Asvadi2023, wang2023age, Maatouk2020}. 
Besides, in \cite{yavascan2021analysis,ahmetoglu2022mista,Ngo2023dec}, novel upgrades to existing random MA techniques have been proposed to further reduce the achieved AoI.  

Recently, the investigations on the application of grant-free random MA techniques to reduce AoI are emerging. Note that in conventional {grant-based MA} schemes, devices need to {carry out hand-shake signaling} scheduling permissions from the base station before transmitting data, resulting in high system overhead
and {latency, particularly in short-packet transmission scenarios}. In contrast, grant-free random MA schemes enable devices to transmit data without prior scheduling requests, thereby reducing
the system overhead and waiting time before transmission.  It {has been shown in} the literature that grant-free schemes have great potential in reducing AoI. In particular,  \cite{Munari21,Huang23,Yu2021} analyzed the performance of AoI achieved by various grant-free access protocols and revealed the key factors which influence AoI. Besides,  \cite{Wang2024age} and \cite{Huang2021age}  proposed novel grant-free transmission strategies to significantly reduce the AoI.
However, the {aforementioned grant-free schemes in} \cite{Munari21,Huang23,Yu2021,Wang2024age,Huang2021age} are mainly based on OMA, where only one device can transmit signal in a single channel resource block, which suffers from
elevated collision probabilities as the traffic load of the system increases. To this end, NOMA assisted grant-free schemes {has been} proposed, where multiple devices can transmit signals simultaneously by sharing the same channel resource block, which can significantly reduce the AoI. 
In \cite{Farhat24,zhang2022IET,Pereira25}, mean-field game theory and deep reinforcement learning methods were introduced to NOMA assisted grant-free transmissions, aiming to minimize AoI 
under the inherent energy constraints. Additionally, in \cite{DingZhiguo2024} and \cite{zhang2021age}, the impact of NOMA-assisted grant-free access on AoI reduction was theoretically analyzed, indicating that NOMA based grant-free schemes can significantly reduce {the average AoI, compared} to  OMA based grant-free schemes. Moreover, in \cite{Ding2024snr}, the AoI of NOMA-assisted grant-free transmission was further enhanced through pre-configured SNR levels.

However, {in order to facilitate the AoI analysis,} most of existing studies on NOMA-assisted grant-free transmissions {adopted} the generate-at-will (GAW) model, which assumes that status updates can be generated and transmitted at any time. While simplifying analysis, this idealized assumption neglects the inherent randomness of update arrivals in many real-world scenarios{, such as} scenarios with event-driven sensing, resulting in a significant mismatch between theoretical results and practical applications. 
According to the best of our knowledge, the rigorous theoretical analysis for the average AoI achieved by NOMA-assisted 
grant-free transmission where status updates are generated randomly is still open. To fulfill the knowledge gap mentioned above, this paper aims to reveal the potential of NOMA-assisted grant-free scheme to reduce AoI by taking into account the randomness of the status update process. The main contributions are listed below.  

\begin{itemize}
    \item Different from existing studies which adopt {the} idealized GAW model to characterize the generation of the status updating packets, a more
    general model is considered in this paper by modeling the randomness of the generation of the status updating packets with {a Bernoulli process}. 
    A novel NOMA-assisted grant-free scheme is adopted, {by pre-configuring} received signal-to-noise ratio (SNR) levels and {carrying} out successive interference cancellation (SIC) to accommodate more simultaneous transmissions in a single time slot. In addition, both {the} transmission strategies  with retransmission and without retransmission are taken into consideration. 
    \item  It is worth pointing out that the derivations of the average AoIs in this paper are much more challenging compared to the commonly adopted GAW model. For example, in the considered retransmission scheme, the status updating processes of different sources are correlated with each other both spatially and temporally. Specifically, on the one hand, the status updatings for different time slots of an individual source are correlated. On the other hand, whether the transmission of a source in a time slot succeeds depends on whether other sources have packets in their buffers and whether they {choose} to transmit. Thorough rigorous derivation, an analytical framework is established to precisely evaluate the average AoIs achieved by the proposed NOMA-assisted grant-free schemes.
   \item Extensive simulation results are provided to verify the accuracy of the developed analytical results.  The impacts of important system parameters on the average AoI are also investigated. It is shown that the average AoIs achieved by NOMA-assisted grant-free schemes are much lower compared to OMA based schemes. It is also shown that the schemes with retransmission  cannot always outperform the schemes without retransmission, except the case where  the status update generation rate is low or the user density is relatively high.
\end{itemize}

\section{System model}\label{system model}
\begin{figure}[!t]
  \centering
    \setlength{\abovecaptionskip}{0em}  
    \setlength{\belowcaptionskip}{-1.5em}   
  \includegraphics[width=3.0in]{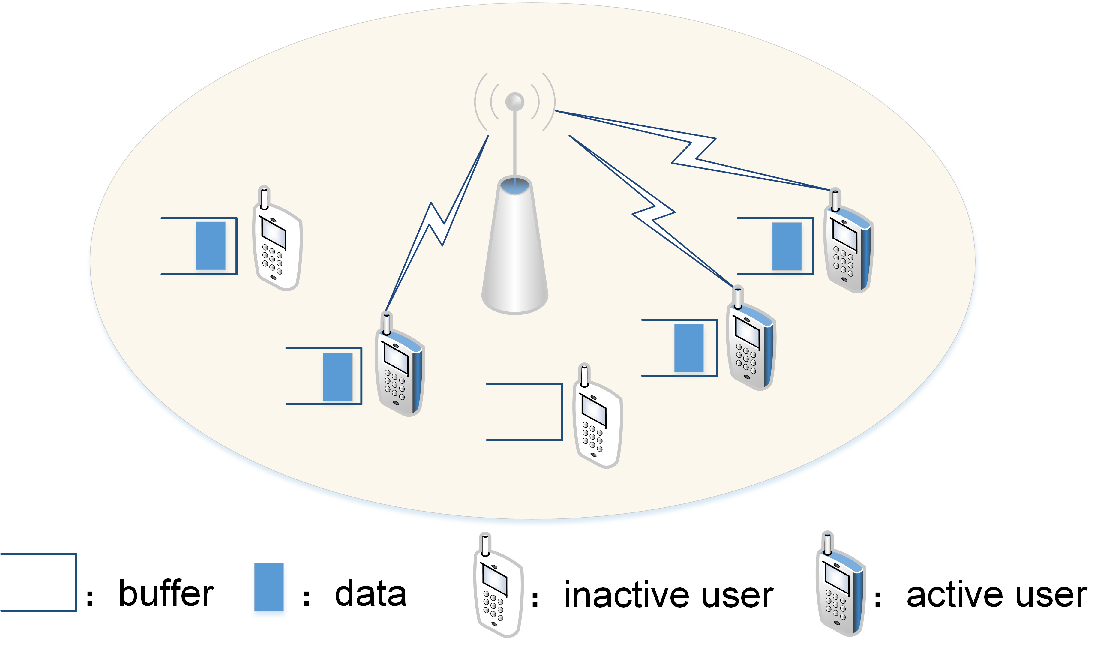}\\
  \caption{Illustration of the system model.}\label{communicationsscenario}
\end{figure}
Consider an uplink random access network, where $M$ source nodes, denoted by $U_m$, $1 \leq m \leq M$, transmit status updates to {the same} destination, as shown in Fig. \ref{communicationsscenario}. {Consider a time framework which consists of multiple time slots,} each with duration $T$.
Bernoulli process is used to model the status updating process of the sources.
Particularly, at each node, status updates are generated at the end of each slot with probability $\lambda$.\footnote{The developed analytical framework by considering Bernoulli status updating process can also apply for Poisson status updating process just with a slight modification.}
Each node maintains a buffer {whose size is assumed to be one}, which means that an older status update packet will be replaced if a newer packet arrives. The buffer will be cleared if the stored packet is successfully transmitted.
\subsection{NOMA assisted grant-free transmission}
In conventional OMA based random access schemes, only {a} single source is admitted to transmit in {one} time slot. In other words, if the number of active sources is larger than one, collision {happens}, which results in transmission failure. Different from conventional OMA based schemes, this paper considers a NOMA based grant-free scheme, {in order to} reduce the collision probability. Specifically, $K$ received SNR levels, denoted by $P_1>P_2>...>P_K$, are pre-configured to be chosen by  {potentially} active sources. The {choices} of the received SNR levels will be given later. At the beginning of each slot, if a source node has a status update packet in its buffer, it makes an attempt to be active (i.e., transmit the packet) with probability $P_{\text{TX}}$ and  inactive with probability $1-P_{\text{TX}}$. If the source has made an attempt to transmit the packet, it will select a received SNR level $P_k$ with probability $q_k$. Note that the required transmit power for $U_m$ to achieve received SNR level $P_k$ is given by $P_k/|h_m^i|^2$, where $h_{m}^i$ is the channel gain of $U_m$ in the $i$-th slot and is modeled as a circularly-symmetric complex Gaussian (CSCG) random variable with zero mean and unit variance. In  practice, it is noteworthy that the required power $P_k/|h_m^i|^2$ might {exceed} the transmitter's maximum power budget $P$. In that case, the source will {choose} to be inactive in this time slot. As a result, the actual probability to be active degrades to:
\begin{align}\label{barptxstart}
\bar{P}_{\text{TX}}
&=P_{\text{TX}}\sum_{k=1}^{K}q_k\left(1-\text{Pr}\left(P_k/|h_m^i|^2>P\right)\right)\\\notag
&=P_{\text{TX}}\sum_{k=1}^{K}q_ke^{-\frac{P_k}{P}},
\end{align}
and if the source is actually activated, the conditional probability for choosing {the} received SNR level $P_k$ becomes
\begin{align}
\bar{q}_k&=P_{\text{TX}}q_k\left(1-\text{Pr}\left(P_k/|h_m^i|^2>P\right)\right)/\bar{P}_{\text{TX}}\\\notag
&=\frac{q_ke^{-\frac{P_k}{P}}}{\sum_{k=1}^{K}q_ke^{-\frac{P_k}{P}}}.
\end{align}

At the destination, successive interference cancellation (SIC) is carried out to decode the transmitted status update packets in an ascending order, i.e., the information at SNR level $P_k$ is decoded prior to that at $P_{k+1}$. To guarantee the success the of SIC, the received SNR levels are set according to the following rules \cite{Ding2024snr}:
\begin{align}\label{P_k}
  \log (1+P_K)=R,
\end{align}
and
\begin{align}\label{log1}
\log (1+\frac{P_k}{1+(M-1)P_{k+1}})=R, 1\leq k\leq (K-1),
\end{align}
where the background noise power is normalized. In addition, $R$ is the minimum {data rate} required to successfully deliver a status update, and it is assumed to be the same for all sources in this paper. Based on the above discussions, source $U_m$ can make
a status update of success in the $i$ -th time interval, if and only if the following four conditions simultaneously hold:
\begin{itemize}
  \item $\mathcal{C}_1$: $U_m$ has a packet in its buffer and {also makes} an attempt for transmission;
  \item $\mathcal{C}_2$: $U_m$ selects received SNR level $P_k$, which is affordable for the power budget $P$;
  \item $\mathcal{C}_3$: No other users also select received SNR level $P_k$, which {leads to a} collision;
  \item $\mathcal{C}_4$: There are no collisions among users which select {those received SNR levels higher than} $P_k$.
\end{itemize}

\subsection{Retransmission strategies}
In this paper, the following two retransmission strategies are considered.
\subsubsection{Without retransmission}
at the end of each transmission slot, the buffer will be emptied regardless of whether the stored packet {has been} transmitted successfully.
\subsubsection{With retransmission}
at the end of each transmission slot, if the transmission of an updating packet is failed, the packet will be kept in the buffer for the coming transmission opportunities, unless new updating packet arrives.

For notational simplification, the considered grant-free NOMA schemes with and without retransmissions are termed ``NOMA-RT'' and ``NOMA-NRT'', respectively, in the rest of the paper.
\subsection{Performance metric}
\begin{figure}[!t]
  \centering
    \setlength{\abovecaptionskip}{0em}  
    \setlength{\belowcaptionskip}{-2em}   
  \includegraphics[width=3in]{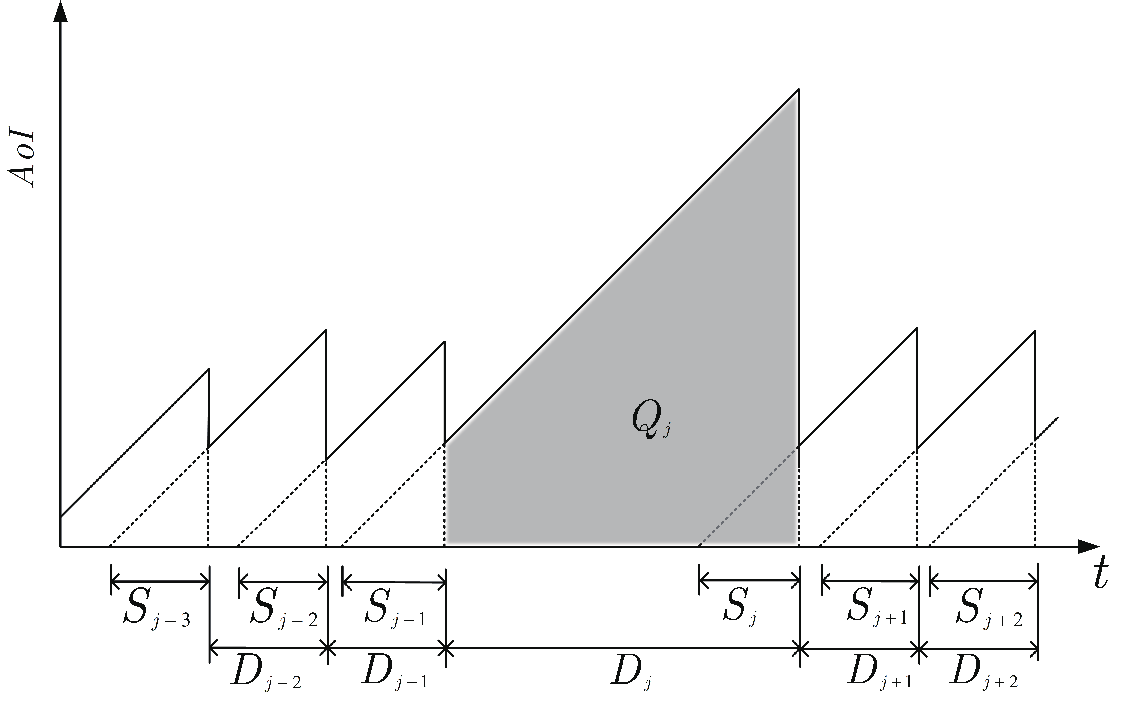}\\
  \caption{Illustration of the AoI of an status updating process.}\label{serration_exp}
\end{figure}

In this paper, the AoI is utilized to quantify the freshness of updates successfully delivered to the receiver \cite{sun2024age}. It is worth noting that only {the updates that are correctly received} are considered relevant to AoI. For a specific source $U_m$, consider $t_j$ as the generation time of the $j$-th status update packet that was successfully transmitted, and let $t_j'$ represent the time it was received at the destination. The instantaneous AoI {is denoted by $\Delta_{m}(t)$, and defined as the} dynamic function that measures the time elapsed since the most recent update available at the receiver was generated. Let $\Omega(t)=\max\{t_j|t_j'\leq t\}$ denote the index of the latest successfully received update observed at the receiver by time $t$. With this definition, the AoI for source $U_m$ is expressed as:
\begin{align}
 \Delta_{m}(t)=t-u(t),
\end{align}
where $u(t)=t_{\Omega(t)}$ denotes the generation timestamp of the latest status update successfully received and observed.
It should be highlighted that the age process $\Delta_m(t)$ exhibits a sawtooth-like pattern \cite{sun2024age}, as shown in Fig. \ref{serration_exp}.

The average AoI for $U_m$ is defined as the time-averaged value of the AoI, {and} is given by:
\begin{align}\label{AAoI}
 \bar{\Delta}_{m}=\lim_{\mathcal{T}\rightarrow \infty} \frac{1}{\mathcal{T}}\int_{0}^{\mathcal{T}}\Delta_{m}(t)\,dt.
\end{align}
The calculation of $\bar{\Delta}_{m}$ proceeds as follows. For simplicity, the interval between the $(j-1)$-th and $j$-th successful deliveries is represented by $D_j=t_j'-t_{j-1}'$, and the system time for a successfully transmitted update is denoted {by} $S_j=t_j'-t_j${, as shown in Fig. \ref{serration_exp}}. Evidently, the computation of the average AoI involves summing a series of trapezoidal areas, called $Q_j$. From Fig. \ref{serration_exp}, it is obtained:
\begin{align}\label{Q_j}
Q_j=D_jS_{j-1}+\frac{1}{2}(D_j)^2.
\end{align}
Therefore, the average AoI $\bar{\Delta}_m$ is given by\cite{yates2021age}:
\begin{align}
\bar{\Delta}_m=\lim_{J\rightarrow \infty}\frac{\sum_{j=1}^{J}{Q_j}}{\sum_{j=1}^{J}{D_j}}.
\end{align}

Moreover, {if} $(D_j,S_j)$ is a stationary and ergodic process, the calculation of $\bar{\Delta}_m$  {can be simplified} to:
\begin{align}\label{exp_AoI}
\bar{\Delta}_m=\frac{\mathbb{E}\{Q_j\}}{\mathbb{E}\{D_j\}}=\frac{\mathbb{E}\{D_jS_{j-1}\}+\mathbb{E}\{(D_j)^2\}/2}{\mathbb{E}\{D_j\}}.
\end{align}

\section{Analysis on AoI for NOMA-NRT and NOMA-RT}\label{NOMA}
In this section, the average AoIs achieved by grant-free NOMA-NRT and grant-free NOMA-RT schemes are analyzed, respectively. Due to symmetry, the average AoIs for different sources are the same if the network is in stable state. Thus, it is sufficient to focus on the average AoI performance of a typical user.
{Therefore, without loss of generality,} the AoI achieved by $U_1$ is focused on in the rest of the paper.

\subsection{AoI achieved by grant-free NOMA-NRT scheme}
As aforementioned in Section II. B,  the buffer of $U_1$ will be emptied regardless of whether the stored
packet {was} transmitted successfully at the end of each time slot. Hence, the results of status updates for different time slots 
of $U_1$ are independent of each other. Therefore, the key for evaluating the average AoI of $U_1$ in the considered grant-free NOMA-NRT scheme
is to obtain the  probability of the event that $U_1$ can complete a successful status update in any given time slot, as characterized in the following lemma. 

\begin{Lemma}\label{lemma1}
In the considered grant-free NOMA-NRT scheme, the probability of the event that $U_1$ can complete a successful status update in any given time slot, denoted by $P_1$,  can be expressed as follows:
\begin{align}\label{Pmall}
P_1=\sum_{i=1}^{M}\sum_{x=1}^{\Gamma}\binom{M-1}{i-1}(\lambda \bar{P}_{\text{TX}})^i(1-\lambda \bar{P}_{\text{TX}})^{M-i}\beta_{(i,x)},
\end{align}
where $\beta_{(i, x)}$ can be expressed as shown in (\ref{beta}),
\begin{figure*}[!t]
\begin{align}
\beta_{(i, x)}=
\begin{cases}
0, &i=x+1, x\ge1 \\
\sum_{k=1}^{K}\bar{q}_k((1\!-\!\sum_{r=1}^{k}\bar{q}_r)^{i\!-\!1}\!-\!\sum_{b=k+1}^{K}(i\!-\!1)\bar{q}_b(1\!-\!\sum_{l=1}^{b}\bar{q}_l)^{i-2}), &x=1, 1\leq i\leq M, i\neq x+1\\
\sum_{k_1=1}^{K-x+1}\sum_{k_2=k_1+x-1}^{K}\binom{i-1}{x-1}x!\bar{q}_{k_1}\bar{q}_{k_2}\\
(\sum_{n_1=k_1+1}^{k_2-x+2}\sum_{n_2=n_1+1}^{k_2-x+3}\cdots\sum_{n_{(x-2)}=n_{(x-3)}+1}^{k_2-1}\bar{q}_{n_1}\bar{q}_{n_2}\cdots \bar{q}_{n_{(x-2)}})\\
((1-\sum_{z=1}^{k_2}\bar{q}_z)^{i-x}-\sum_{n=k_2+1}^{K}(i-x)\bar{q}_n(1-\sum_{g=1}^{n}\bar{q}_g)^{i-x-1}),  &2\leq x\leq \Gamma, 1\leq i\leq M, i\neq x+1.
\end{cases}
\label{beta}
\end{align}
\end{figure*}
$\Gamma=\mathds{1}_{K}(i)i+(1-\mathds{1}_{K}(i))(K-1)$, {and} $\mathds{1}_{K}(i)$ is an indicator function so that $\mathds{1}_{K}(i)=1$ if $i\leq K$ and  $\mathds{1}_{K}(i)=0$ otherwise.
\end{Lemma}

\begin{IEEEproof}
By noting that the status updating process of $U_1$ is affected by other users which are {also} actually active in the considered slot, $P_1$ can be written as:
\begin{align}\label{Pm}
P_1=\sum_{i=1}^{M}\sum_{x=1}^{\Gamma}P\{X=x|I=i\}P\{I=i\},
\end{align}
where $I$ denotes the number of actually active users in the considered time slot which {is also used} by $U_1$,
and $X$  is the number of users whose status updates are successfully transmitted among the $I$ active users. Note that $U_1$ is also included in $X$.
Additionally, it is important to note that the number of actually active users is not necessarily equal to the maximum number of successful users, as it is limited by the number of SNR levels. Specifically, when $i>K$, it means that at least one SNR level is occupied and in collision, so the maximum number of successful users can only be $K-1$. On the other hand, {for the case of} $i\leq K$, the maximum number of successful users can be up to $i$. Therefore, it can be concluded that $1\leq x\leq \Gamma$.

By noting that the conditions $\mathcal{C}_1$ and $\mathcal{C}_2$  described in section II must be satisfied to {ensure} that a source is actually active, thus, the expression for $P\{I=i\}$ can be easily obtained as follows:
\begin{align}\label{pIi}
P\{I=i\}=\binom{M-1}{i-1}(\lambda \bar{P}_{\text{TX}})^i(1-\lambda \bar{P}_{\text{TX}})^{M-i}.
\end{align}

The remaining task is to evaluate $P\{X=x|I=i\}$, $1\leq x\leq \Gamma$, $1\leq i\leq M$.
For notational simplicity, {define} $\beta(i,x)=P\{X=x|I=i\}$.
Firstly, it can be observed that $\beta(x+1,x)=0$, since as long as a collision
occurs, the transmissions of at least two sources will fail. For other values of $X$, the evaluation of $\beta(i,x)$
can be classified into the following two cases:

\subsubsection{{The case with} $x=1$, $1\leq i\leq M$, $i\neq x+1$}
for this case, the conditions which {cause} $X=1$ are as follows: a) $U_1$ selects a received SNR level $P_k$; b) all other active users select SNR levels lower than $P_k$ and  none of them succeed.
Hence, $\beta(i,x)$ can be expressed as follows:
\begin{align}\label{s3}
&\beta(i,x)\\\notag
=&\sum_{k=1}^{K}\bar{q}_k((1\!-\!\sum_{r=1}^{k}\bar{q}_r)^{i-1}\!-\!\sum_{b=k+1}^{K}(i\!-\!1)\bar{q}_b(1\!-\!\sum_{l=1}^{b}\bar{q}_l)^{i-2}).
\end{align}

\subsubsection{{The case with} $2\leq x\leq \Gamma$, $1\leq i\leq M$, $i\neq x+1$}
for this case, the conditions which {cause} $X=x$ are as follows: a) there are $x$ sources ($U_1$ is included) which select distinct SNR levels, among which the maximum and minimum are denoted by $P_{k_1}$ and $P_{k_2}$ ($1\leq k_1\leq k_2\leq K$), respectively;
b) except the $x$ sources,  all other active users select {the} SNR levels lower than $P_{k_2}$ and none of them succeed.
Based on the above two conditions, $P\{X=x|I=i, 2\leq x\leq \Gamma, 1\leq i\leq M, i\neq x+1\}$ can be expressed as follows:
\begin{align}\label{s4}
\beta(i,x)
&=\sum_{k_1=1}^{K-x+1}\sum_{k_2=k_1+x-1}^{K}\binom{i-1}{x-1}x!\bar{q}_{k_1}\bar{q}_{k_2}\\\notag
&(\sum_{n_1=k_1+1}^{k_2-x+2}\sum_{n_2=n_1+1}^{k_2-x+3}\cdots\sum_{n_{(x-2)}=n_{(x-3)}+1}^{k_2-1}\bar{q}_{n_1}\bar{q}_{n_2}\cdots \bar{q}_{n_{(x-2)}})\\\notag
&((1-\sum_{z=1}^{k_2}\bar{q}_z)^{i-x}-\sum_{n=k_2+1}^{K}(i-x)\bar{q}_n(1-\sum_{g=1}^{n}\bar{q}_g)^{i-x-1}).
\end{align}

By taking (\ref{pIi})-(\ref{s4}) into (\ref{Pm}), the expression for $P_1$ can be obtained. Therefore, the proof is complete.
\end{IEEEproof}

Applying Lemma $1$, the expression for the average AoI achieved by the considered grant-free NOMA-NRT scheme can be obtained as highlighted in the following theorem.

\begin{Theorem}\label{theorem1}
The average AoI achieved by the considered grant-free NOMA-NRT scheme for $U_1$, denoted by $\bar{\Delta}_{1}^{NOMA-NRT}$, can be expressed as follows:
\begin{align}\label{pmnew}
 &\bar{\Delta}_{1}^{NOMA-NRT}=\frac{T}{2}+\frac{T}{P_1},
\end{align}
\end{Theorem}

\begin{IEEEproof}
For the considered grant-free NOMA-NRT scheme, it can be easily found that $D_j$ and $S_{j-1}$ are independent with each other. Thus, the expression for $\bar{\Delta}_1^{NOMA-NRT}$ shown in (\ref{exp_AoI}) can be simplified  as follows:
\begin{align}\label{}
\bar{\Delta}_1^{NOMA-NRT}&=\mathbb{E}\{S_{j-1}\}+\frac{\mathbb{E}\{D_j^2\}/2}{\mathbb{E}\{D_j\}}\\\notag
&=T+\frac{\mathbb{E}\{D_j^2\}/2}{\mathbb{E}\{D_j\}},
\end{align}
where the last step follows from the fact that $\mathbb{E}\{S_{j-1}\}=T$.
Thus, the remaining task is to calculate $\mathbb{E}\{D_j\}$ and $\mathbb{E}\{D_j^2\}$.

Note that the value of $D_j$ can be expressed as $D_j=NT$, where $N$ is a random positive integer following geometric distribution. Thus, it can be straightforwardly obtained that
\begin{align}
P(D_j=N T)=P_1(1-P_1)^{N-1}.
\end{align}

Thus,  the expressions for $\mathbb{E}\{D_j\}$ and $\mathbb{E}\{D_j^2\}$  can be easily obtained as follows:
\begin{align}\label{eqD}
\mathbb{E}\{D_j\}=\sum_{N=1}^{\infty}(NT)P(D_j=NT)=\frac{T}{P_1},
\end{align}
\begin{align}\label{eqD2}
\mathbb{E}\{D_j^2\}=\sum_{N=1}^{\infty}(NT)^2P(D_j=NT)=\frac{T^2(2-P_1)}{P_1^2}.
\end{align}
The proof is complete.
\end{IEEEproof}

Based on (\ref{pmnew}), it is interesting to optimize to transmission probability $P_{\text{TX}}$ for minimizing $\bar{\Delta}_1^{NOMA-NRT}$. From the results shown in Theorem \ref{theorem1}, it is obvious that the minimization of $\bar{\Delta}_1^{NOMA-NRT}$ is equivalent to the maximization of $P_1$. Thus, it is necessary to investigate the monotonicity of $P_1$. To this end, let's first rewrite the expression of $P_1$ shown in (\ref{Pmall}) as follows:
\begin{align}\label{}
P_1=&\sum_{i=1}^{K}\sum_{x=1}^{i}\binom{M\!-\!1}{i\!-\!1}(\lambda \bar{P}_{\text{TX}})^i(1-\lambda \bar{P}_{\text{TX}})^{M-i}\beta_{(i,x)}\\\notag
&+\sum_{i=K+1}^{M}\sum_{x=1}^{K-1}\binom{M\!-\!1}{i\!-\!1}(\lambda \bar{P}_{\text{TX}})^i(1-\lambda \bar{P}_{\text{TX}})^{M-i}\beta_{(i,x)}.
\end{align}

Then, with some straightforward manipulations, the derivative of $P_1$ with respect to $\bar{P}_{\text{TX}}$ can be
obtained as follows:
\begin{align}\label{P1de}
P_1'=&\sum_{i=1}^{K}\sum_{x=1}^{i}\binom{M-1}{i-1}\lambda(\lambda \bar{P}_{\text{TX}})^{i-1}(1-\lambda \bar{P}_{\text{TX}})^{M-i-1}\\\notag
&(i-\lambda \bar{P}_{\text{TX}}M)\beta_{(i,x)}+\sum_{i=K+1}^{M}\sum_{x=1}^{K-1}\binom{M-1}{i-1}\lambda\\\notag
&(\lambda \bar{P}_{\text{TX}})^{i-1}(1-\lambda \bar{P}_{\text{TX}})^{M-i-1}(i-\lambda \bar{P}_{\text{TX}}M)\beta_{(i,x)}
\end{align}
Unfortunately, it is complicated to characterize the properties of $P_1'$  from (\ref{P1de}) for general choices of values of $K$. Therefore, the following will focus on the special case of $K=2$ for {optimizing $P_{\text{TX}}$ to obtain} some insights.

\begin{Corollary}
When $K=2$ and $M\to \infty$, the
optimal $P_{\text{TX}}$ for minimizing $\bar{\Delta}_1^{NOMA-NRT}$ can be approximated as follows:
\begin{align}
P_{\text{TX}}^*\approx\frac{\eta}{\lambda Mq_1(q_1e^{\frac{-(2^R-1)2^R}{P}}+q_2e^{{\frac{-(2^R-1)}{P}})}}, 
\end{align}
where $\eta$ is the solution of the following equation: $(\frac{-\eta^2q_2}{q_1}+\frac{(2q_1-1)q_2\eta}{q_1}+q_2)e^{\frac{-\eta}{q_1}}+(1-\eta)q_1 e^{-\eta}=0$, {with} $\frac{\sqrt{1+4q_1^2}}{2}\leq \eta\leq 1$.
\end{Corollary}
\begin{IEEEproof}
The proof will first obtain $\bar{P}_{\text{TX}}^*$, based on which $P_{\text{TX}}^*$ can be obtained according to (\ref{barptxstart}). 
When $K=2$, $P_1'$ can be explicitly expressed as shown in (\ref{K2P1de}) at the top of next page.
\begin{figure*}[!t]
\begin{align}\label{K2P1de}
P_1'=&\lambda(1\!-\!\lambda \bar{P}_{\text{TX}})^{M\!-\!2}(1\!-\!\lambda \bar{P}_{\text{TX}}M)+(M\!-\!1)\lambda(\lambda \bar{P}_{\text{TX}})(1-\lambda \bar{P}_{\text{TX}})^{M-3}(2-\lambda \bar{P}_{\text{TX}}M)2\bar{q}_1\bar{q}_2+\\\notag
&
\sum_{i=3}^{M}\binom{M-1}{i-1}\lambda
(\lambda \bar{P}_{\text{TX}})^{i-1}(1-\lambda \bar{P}_{\text{TX}})^{M-i-1}(i-\lambda \bar{P}_{\text{TX}}M)\bar{q}_1\bar{q}_2^{i-1}\\\notag
=&\underbrace{((1-(M-1)\bar{q}_{k1})\bar{q}_{k2}M\lambda^3\bar{P}_{\text{TX}}^2+(2\bar{q}_{k1}(M-1)-1-M)q_2\lambda^2\bar{P}_{\text{TX}}+\lambda\bar{q}_{k2})(1-\lambda\bar{P}_{\text{TX}})^{M-3}}_{f_1}\\\notag
&+\underbrace{(1-M\bar{q}_{k1}\lambda\bar{P}_{\text{TX}})\bar{q}_{k1}\lambda(1-\bar{q}_{k1}\lambda\bar{P}_{\text{TX}})^{M-2}}_{f_2}
\end{align}
\end{figure*}
As shown in (\ref{K2P1de}), $P_1'$ can be divided into two parts, namely $f_1$ and $f_2$, respectively. The signs of $f_1$ and $f_2$ can be determined as {follows}: 

    a) For  $f_{1}$,  it is {straightforward to verify that the roots of $f_{1}=0$ can} be expressed as:
      \begin{align}\label{root_f11}
\frac{1\!-\!2q_1(M\!-\!1)\!+\!M\!\pm\!\sqrt{(1\!-\!M)(1\!+\!4q_1\!+\!4q_1^2\!-\!M\!-\!4Mq_1^2)}}{2M\lambda(1\!+\!q_1\!-\!q_1M)}.
\end{align}
When $M\to\infty$,  the expressions shown in Eq. (\ref{root_f11}) can be simplified  to  $\pm\frac{\sqrt{1+4q_1^2}}{2\lambda Mq_1}$.
Further, it can be proved that $f_{1}\ge0$ when  $-\frac{\sqrt{1+4q_1^2}}{2\lambda Mq_1}\leq\bar{P}_{\text{TX}}\leq\frac{\sqrt{1+4q_1^2}}{2\lambda Mq_1}$, and $f_{1}<0$ otherwise. 

b) For $f_{2}$, it is straightforward to obtain that $f_{2}\geq 0$ when $\bar{P}_{\text{TX}}\leq \frac{1}{\lambda Mq_1}$, and $f_2<0$ otherwise.

From the aforementioned observations, it can be proved that:  
a) when $0< \bar{P}_{\text{TX}}<\frac{\sqrt{1+4q_1^2}}{2\lambda Mq_1}$, $P'_1>0$;  b) when $\frac{1}{\lambda Mq_1}<\bar{P}_{\text{TX}}<1$, $P'_1<0$.
Therefore, 
there is at least one root for $P'_1=0$, which is exactly $\bar{P}_{\text{TX}}^*$ {with the following range}:  
\begin{align}\label{P_TX_range}
\frac{\sqrt{1+4q_1^2}}{2\lambda Mq_1}\leq \bar{P}_{\text{TX}}^*\leq \frac{1}{\lambda Mq_1}.
\end{align}
From (\ref{P_TX_range}), it can be easily observed that $\bar{P}_{\text{TX}}^*$ can be denoted by
$\bar{P}_{\text{TX}}^*=\frac{\eta}{\lambda Mq_1}$, 
where $\eta$ is a parameter to be determined and $\frac{\sqrt{1+4q_1^2}}{2}\leq \eta\leq 1$. Then,   $P_1'$ can be rewritten as follows:
\begin{align}\label{barfptx}
P_1'=&(\frac{-\lambda\eta^2q_2}{q_1}\!+\!\frac{(2q_1\!-\!1)q_2\lambda\eta}{q_1}\!+\!\lambda q_2)(1\!-\!\frac{\eta}{Mq_1})^{M\!-\!3}+\\\notag
&(1-\eta)q_1\lambda(1-\frac{\eta}{M})^{M-2}\\\notag
\approx&(\frac{-\lambda\eta^2q_2}{q_1}\!+\!\frac{(2q_1\!-\!1)q_2\lambda\eta}{q_1}\!+\!\lambda q_2)e^{\frac{-\eta}{q_1}}+(1\!-\!\eta)q_1\lambda e^{-\eta}\\\notag
\end{align}
where the last step follows from the fact that $\lim\limits_{M\to \infty}\ln(1+\frac{\eta}{M})^M=\eta$.
Then,  $\eta$ can be obtained by solving the equation $P'_1=0$.  Therefore, the expression for $\bar{P}_{\text{TX}}^*$ can be {obtained}. Finally,  the optimal choice of $P_{\text{TX}}$ can be obtained according to (\ref{barptxstart}) and the proof is complete.
\end{IEEEproof}
\begin{Remark}
From Corollary $1$, it is obvious that when $M$ is sufficiently large, $P_{\text{TX}}^*$ should become smaller as $\lambda$ and $M$ increases. This observation is consistent with the intuition that the collision probability will increase as $\lambda$ and $M$ increase, so that 
a smaller transmission probability is necessary to reduce collision probability. 
\end{Remark}

\subsection{AoI achieved by grant-free NOMA-RT scheme}
Due to the adoption of the considered re-transmission strategy, the status updatings for different time slots of an individual source {become} correlated. Moreover, whether the transmission of a source in a time slot succeeds depends on whether other sources have packets in their 
buffers and whether they select to transmit. Therefore, the status updating processes of different sources are coupled with each other, making the 
analysis for the average AoI of the grant-free NOMA-RT scheme challenging. To deal with the aforementioned {challenges}, it is useful to first 
characterize the probability of a successful status updating of a source in a time slot when the whole status updating system is in stable state, as highlighted in the following lemma.  

\begin{Lemma}\label{lemma2}
In the considered grant-free NOMA-RT scheme, given that there is packet to be transmitted in $U_1$'s buffer at the beginning of a slot, the probability that $U_1$ selects to transmit and transmits successfully,
which is denoted by $\tilde{P_1}$, can be expressed as follows:
\begin{align}\label{Psall}
\tilde{P_1}=&\sum_{m=0}^{M-1}\sum_{x=1}^{\min\{m+1,K\}}\sum_{i=x}^{\mathds{1}_{K}(x)K+(1-\mathds{1}_{K}(x))(m+1)}\binom{m}{i\!-\!1}\\\notag
&(1-\bar{P}_{\text{TX}})^{m-i+1}\bar{P}_{\text{TX}}^i\beta_{(i, x)}\frac{ \frac{\binom{M-1}{m}}{\binom{M}{m+1}}\pi_{m+1}}{\sum_{m=0}^{M-1}\frac{\binom{M-1}{m}}{\binom{M}{m+1}}\pi_{m+1}},
\end{align}
where $\mathds{1}_{K}(x)=1$ if $x==K$, otherwise $\mathds{1}_{K}(x)=0$, and $\pi_{m}$ ($1\leq m\leq M$) is the solution of the following equations:
\begin{align}\label{pim}
&\begin{bmatrix}
\pi_0 & \pi_1 &\dots &\pi_M
\end{bmatrix}
\begin{bmatrix}
P_{0\to 0} & P_{0\to 1} & \dots & P_{0\to M}\\
P_{1\to 0} & P_{1\to 1} & \dots & P_{1\to M}\\
\vdots &  \vdots & \vdots & \vdots\\
P_{M\to 0} & \dots &\dots & P_{M\to M}\\
\end{bmatrix}\\\notag
=&
\begin{bmatrix}
\pi_0 & \pi_1 &\dots &\pi_M
\end{bmatrix}.
\end{align}
with
\begin{align}\label{a6}
P_{b\to a}=&\sum_{x^*=\max\{b-a,0\}}^{\min\{b,K\}}\binom{M-b+x^*}{a-b+x^*}\lambda^{a-b+x^*}\\\notag
&(1-\lambda)^{M-a}\sum_{i^*=x^*}^{\mathds{1}_{K}(x^*)K+(1-\mathds{1}_{K}(x^*))b}\binom{b}{i^*}\\\notag
&(1-\bar{P}_{\text{TX}})^{b-i^*}\bar{P}_{\text{TX}}^{i^*}\hat{\beta}_{(i^*, x^*)}, 0\leq a\leq M, 0\leq b\leq M.
\end{align}
where $\mathds{1}_{K}(x^*)=1$ if {$x^*=K$}, otherwise $\mathds{1}_{K}(x^*)=0$, 
 and $\hat{\beta}_{(i^*, x^*)}$ is shown in (\ref{betahat})
\begin{figure*}[h!]
\small
\begin{align}
	\hat{\beta}_{(i^*, x^*)} \!\!=\!\! \begin{cases}
	      0, &i^*=x^*+1, x^*\ge0\\
          1-\sum_{k=1}^{K}i^*\bar{q}_k(1-\sum_{r=1}^{k}\bar{q}_r)^{i^*-1}, &x^*=0, 1\leq i^*\leq M, i^*\neq x^*+1\\
          \sum_{k=1}^{K}i^*\bar{q}_k((1\!\!-\!\sum_{r=1}^{k}\bar{q}_r)^{i^*\!\!-\!1}\!-\!\sum_{b=k+1}^{K}(i^*\!\!-\!1)\bar{q}_b(1\!-\!\sum_{l=1}^{b}\bar{q}_l)^{i^*\!-\!2}), &x^*=1, 1\leq i^*\leq M, i^*\neq x^*+1\\
          \sum_{k_1=1}^{K-x^*+1}\sum_{k_2=k_1+x^*-1}^{K}\binom{i^*}{x^*}x^*!\bar{q}_{k_1}\bar{q}_{k_2}\\
          (\sum_{n_1=k_1+1}^{k_2-x^*+2}\sum_{n_2=n_1+1}^{k_2-x^*+3}\cdots\sum_{n_{(x^*-2)}=n_{(x^*-3)+1}}^{k_2-1}\bar{q}_{n_1}\bar{q}_{n_2}\cdots \bar{q}_{n_{(x^*-2)}})\\
          ((1\!-\!\sum_{z=1}^{k_2}\bar{q}_z)^{i^*\!-\!x^*}\!-\!\sum_{n=k_2+1}^{K}(i^*\!-\!x^*)\bar{q}_n(1\!-\!\sum_{g=1}^{n}\bar{q}_g)^{i^*\!-\!x^*\!-\!1}), &2\leq x^*\leq \tilde{\Gamma}, 1\leq i^*\leq M, i^*\neq x^*+1.
		   \end{cases}
\label{betahat}
\end{align}
\end{figure*}
with $\tilde{\Gamma}=\mathds{1}_{K}(i^*)i^*+(1-\mathds{1}_{K}(i^*))(K-1)$, where $\mathds{1}_{K}(i^*)=1$ if $i^*\leq K$, otherwise $\mathds{1}_{K}(i^*)=0$.
\end{Lemma}
\begin{IEEEproof}
\begin{figure}[t!]
  	\centering
	\vspace{0em}
	\setlength{\abovecaptionskip}{0em}   
	\setlength{\belowcaptionskip}{-1em}
  \includegraphics[width=2.5in]{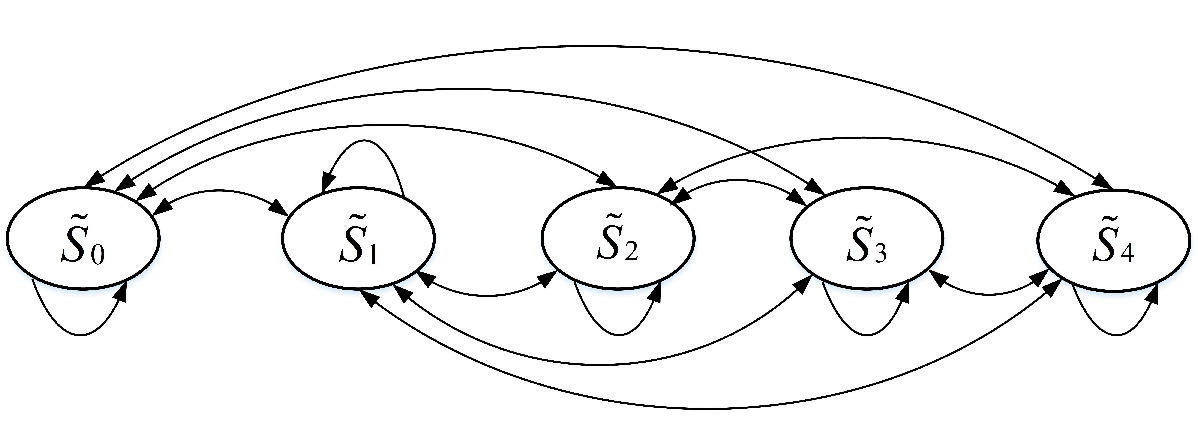}\\
  \caption{ Illustration the process of transferring the number of users with data in steady state. $M=5$.}\label{Markov222}
\end{figure}
By noting that the transmission of $U_1$ is affected by other users which also have data to transmit in their buffers at the beginning of the considered slot, $\tilde{P_1}$ can be expressed as:
\begin{align}\label{Ps}
\tilde{P_1}=&\sum_{m=0}^{M-1}\sum_{x=1}^{\min\{m+1,K\}}P\{X=x|\mathcal{A}_m\}P\{\mathcal{A}_m\},
\end{align}
where $\mathcal{A}_m$ denotes the event that in addition to $U_1$, there are other $m$ users which also have {packets} to transmit in their buffers. Besides, the random variable $X$ denotes the number of users which transmit successfully in the considered time slot.

To obtain $P\{X=x|\mathcal{A}_m\}$, it is necessary to consider the number of users which are actually active, since each user with data may possibly choose to be silent. Denote $I$ {by} the number users which are actually active, where $U_1$ is included, $P\{X=x|\mathcal{A}_m\}$ can be expressed as:
\begin{align}\label{pxb}
P\{X=x|\mathcal{A}_m\}
=&\sum_{i=x}^{\mathds{1}_{K}(x)K+(1-\mathds{1}_{K}(x))(m+1)}\binom{m}{i-1}\\\notag
&(1-\bar{P}_{\text{TX}})^{m-i+1}\bar{P}_{\text{TX}}^iP\{X=x|I=i\},
\end{align}
where $P\{X=x|I=i\}$ is the conditional probability of $X$ given $I$, which has already been obtained as shown in (\ref{beta}). Besides, it is noteworthy that the range of the possible values for $I$ is dependent on $x$. Specifically,
when $x<K$, the value range of $I$ is from $x$ to $m+1$; when $x=K$, $I$ can {only be} $K$.

Then, the remaining task is to evaluate $P\{\mathcal{A}_m\}$. Note that the number of users which have packets in their {buffers} at the beginning of consecutive slots can be modeled as a Markov chain as shown in  Fig.  \ref{Markov222}, where $\tilde{S}_a$, $0\leq a \leq M$, denotes the state {in which} $a$ users have packets in their buffer at the beginning of the considered slot. It is noteworthy that it is {possible that $U_1$ is included $\tilde{S}_a$}.
Note that $P\{\mathcal{A}_m\}$ can be obtained though the stationary state probability of the Markov chain. Thus, it is necessary to calculate the stationary state distribution of the Markov chain, which can be denoted by a vector $\mathbf{\pi}=[\pi_0,\pi_1,\cdots,\pi_M]$, where $\pi_a$ is the stationary probability in state $\tilde{S}_a$.
Then, $\pi_a$ can be obtained by solving the following stationary state system of equations:
\begin{align}\label{pab}
\pi_a=\sum_{b=0}^{M}\pi_bP_{b\to a}, 0\leq a\leq M, 0\leq b\leq M,
\end{align}
where $P_{b\to a}$ denotes the transition probability from state $\tilde{S}_b$ to state $\tilde{S}_a$, which will be evaluated as in the following.

To obtain $P_{b\to a}$, it is necessary to observe that the conditions which yield the transfer from state $\tilde{S}_b$ (at the beginning of the $t-1$-th slot) to state $\tilde{S}_a$  (at the beginning of the $t$-th slot) are {two-fold}:
\begin{itemize}
  \item there are $X^*$ users in the $t-1$-th slot successfully completing their status updating,
where $\max\{b-a,0\}\leq X^*\leq \min\{b,K\}$;
  \item at the end of the $t-1$-th slot,  there are $a-(b-X^*)$ users which have new packets arrive.
\end{itemize}
Therefore, $P_{b\to a}$ can be expressed as follows:
\begin{align}\label{}
P_{b\to a}=&\sum_{x^*=\max\{b-a,0\}}^{\min\{b,K\}}\binom{M-b+x^*}{a-b+x^*}\lambda^{a-b+x^*}\\\notag
&(1-\lambda)^{M-a}P\{X^*=x^*|\tilde{S}_b\}\\\notag
{=}&\sum_{x^*=\max\{b-a,0\}}^{\min\{b,K\}}\binom{M-b+x^*}{a-b+x^*}\lambda^{a-b+x^*}(1-\lambda)^{M-a}\\\notag
&\sum_{i^*=x^*}^{\mathds{1}_{K}(x^*)K+(1-\mathds{1}_{K}(x^*))b}\binom{b}{i^*}(1-\bar{P}_{\text{TX}})^{b-i^*}\bar{P}_{\text{TX}}^{i^*}\\\notag
&{P\{X^*=x^*|I^*=i^*\}},
\end{align}
where $I^*$ denotes the number of users among the $b$ users which are actually active.
Note that the difference between $I^*$, $X^*$ and $I$, $X$ in Lemma $1$ is that $I^*$ and $X^*$ not necessarily
include $U_1$. For notational simplicity, {define}  $\hat{\beta}(i^*,x^*)=P\{X^*=x|I^*=i^*\}$.
Then, similar to the steps for obtaining $\beta(i,x)$,
the calculation for $\hat{\beta}(i^*,x^*)$ can be classified into
the following four cases:

\subsubsection{$i^*=x^*+1$, $x^*\ge0$}
\begin{align}\label{a1}
\hat{\beta}(i^*,x^*)=0.
\end{align}
\subsubsection{$x^*=0$, $1\leq i^*\leq M$, $i^*\neq x^*+1$}
\begin{align}\label{a2}
\hat{\beta}(i^*,x^*)=1-\sum_{k=1}^{K}i^*\bar{q}_k(1-\sum_{r=1}^{k}\bar{q}_r)^{i^*-1}.
\end{align}
\subsubsection{$x^*=1$, $1\leq i^*\leq M$, $i^*\neq x^*+1$}
\begin{align}\label{a3}
&\hat{\beta}(i^*,x^*)\\\notag
=&\sum_{k=1}^{K}i^*\bar{q}_k((1\!-\!\sum_{r=1}^{k}\bar{q}_r)^{i^*\!-\!1}\!-\!\!\!\!\sum_{b=k+1}^{K}(i^*\!-\!1)\bar{q}_b(1\!-\!\sum_{l=1}^{b}\bar{q}_l)^{i^*\!-\!2}).
\end{align}
\subsubsection{$2\leq x^*\leq \tilde{\Gamma}$, $1\leq i^*\leq M$, $i^*\neq x^*+1$}
\begin{align}\label{a4}
&\hat{\beta}(i^*,x^*)\\\notag
&=\sum_{k_1=1}^{K-x^*+1}\sum_{k_2=k_1+x^*-1}^{K}\binom{i^*}{x^*}x^*!\bar{q}_{k_1}\bar{q}_{k_2}\\\notag
&(\sum_{n_1=k_1+1}^{k_2-x^*+2}\sum_{n_2=n_1+1}^{k_2-x^*+3}\cdots\sum_{n_{(x^*-2)}=n_{(x^*-3)+1}}^{k_2-1}\bar{q}_{n_1}\bar{q}_{n_2}\cdots \bar{q}_{n_{(x^*-2)}})\\\notag
&((1-\sum_{z=1}^{k_2}\bar{q}_z)^{i^*-x^*}-\sum_{n=k_2+1}^{K}(i^*-x^*)\bar{q}_n(1-\sum_{g=1}^{n}\bar{q}_g)^{i^*-x^*-1}),
\end{align}

Now, the expression for $P_{b\to a}$ is obtained{, and} hence, $\pi_a$ ($0\leq a\leq M$) can be obtained from (\ref{pab}).

Further, based on $\pi_a$, the expression for $P\{\mathcal{A}_m\}$ can be obtained as in the following.
First, define $\hat{\pi}_{m+1}^1$ as the stationary probability for the Markov chain that there are $m+1$ users which have packets to transmit in their buffers for a time slot where $U_1$ is included. Due to the symmetry among the sources, $\hat{\pi}_{m+1}^1$ can be easily obtained as follows:
\begin{align}\label{pab1}
\hat{\pi}_{m+1}^1=\frac{\binom{M-1}{m}}{\binom{M}{m+1}}\pi_{m+1}.
\end{align}
Then, $P\{\mathcal{A}_m\}$ can be obtained by taking normalization as follows:
\begin{align}\label{pab2}
P\{\mathcal{A}_m\}=\frac{ \hat{\pi}_{m+1}^1}{\sum_{m=0}^{M-1} \hat{\pi}_{m+1}^1}.
\end{align}
Therefore, the proof is complete.
\end{IEEEproof}

Based on Lemma \ref{lemma2}, the expression for the average AoI achieved by the considered grant-free NOMA-RT scheme can be obtained as highlighted in the following theorem.

\begin{Theorem}\label{theorem2}
The average AoI achieved by the considered grant-free NOMA-RT scheme for $U_1$, denoted by $\bar{\Delta}_{1}^{N\!O\!M\!A\!-\!R\!T\!}$, can be expressed as:
\begin{align}\label{NOMART}
\bar{\Delta}_{1}^{N\!O\!M\!A\!-\!R\!T\!}&=\frac{T}{1-(1-\lambda)(1-\tilde{P_1})}+\\\notag
&T(\frac{2(\lambda^2\!+\!\tilde{P_1}^2)\!-\!\lambda \tilde{P_1}(3\lambda\!+\!3\tilde{P_1}\!-\!\lambda\tilde{P_1}\!-\!2)}{2\lambda \tilde{P_1}(\lambda\!+\!\tilde{P_1}\!-\!\lambda\tilde{P_1})}),
\end{align}
\end{Theorem}

\begin{IEEEproof}
It can be easily {shown} that $D_j$ and $S_{j-1}$ are independent {from} each other, thus,
$\bar{\Delta}_1^{NOMA-RT}$ can be expressed as:
\begin{align}\label{}
\bar{\Delta}_1^{NOMA-RT}=\frac{\mathbb{E}\{Q_j\}}{\mathbb{E}\{D_j\}}=\mathbb{E}\{S_{j-1}\}+\frac{\mathbb{E}\{D_j^2\}/2}{\mathbb{E}\{D_j\}}.
\end{align}
\setcounter{subsubsection}{0}
\subsubsection{Evaluation of $\mathbb{E}\{S_{j-1}\}$}
Note that {there is} an implicit condition for $S_{j-1}$ that $U_1$ successfully completes a status {update}, which
is denoted by event $\mathcal{B}$.  Moreover, it can be found that $S_{j-1}$ can be expressed as $S_{j-1}=\tilde{N} T$, where  $\tilde{N}$ is a random positive integer following {the} geometric distribution. Thus, $\mathbb{E}\{S_{j-1}\}$ can be evaluated as follows:
\begin{align}\label{Sj-1}
\mathbb{E}\{S_{j-1}\}
=\sum_{\tilde{N}=1}^{\infty}\frac{(\tilde{N} T)P\{S_{j-1}=\tilde{N} T,\mathcal{B}\}}{P\{\mathcal{B}\}}.
\end{align}
It can be straightforwardly obtained that
\begin{align}\label{PSB}
P(S_{j-1}=\tilde{N}T, \mathcal{B})=\lambda \tilde{P_1}(1-\lambda)^{\tilde{N}-1}(1-\tilde{P_1})^{\tilde{N}-1}.
\end{align}
Hence, the expression for $P\{\mathcal{B}\}$ can be obtained as:
\begin{align}\label{PB}
P\{\mathcal{B}\}=&\sum_{\tilde{N}=1}^{\infty}P(S_{j-1}=\tilde{N} T,\mathcal{B})\\\notag
=&\frac{\lambda \tilde{P_1}}{1-(1-\tilde{P_1})(1-\lambda)}.
\end{align}
By taking (\ref{PSB}) and (\ref{PB}) into (\ref{Sj-1}), the expression for $\mathbb{E}\{S_{j-1}\}$ can be obtained  which can be expressed as follows:
\begin{align}\label{}
\mathbb{E}\{S_{j-1}\}=\frac{T}{1-(1-\tilde{P_1})(1-\lambda)}.
\end{align}

\subsubsection{Evaluation of $\mathbb{E}\{D_j\}$ and $\mathbb{E}\{D_j^2\}$}
\begin{figure}[!t]
  	\centering
	\vspace{0em}
	\setlength{\abovecaptionskip}{0em}   
	\setlength{\belowcaptionskip}{-1em}
  \includegraphics[width=1.8in]{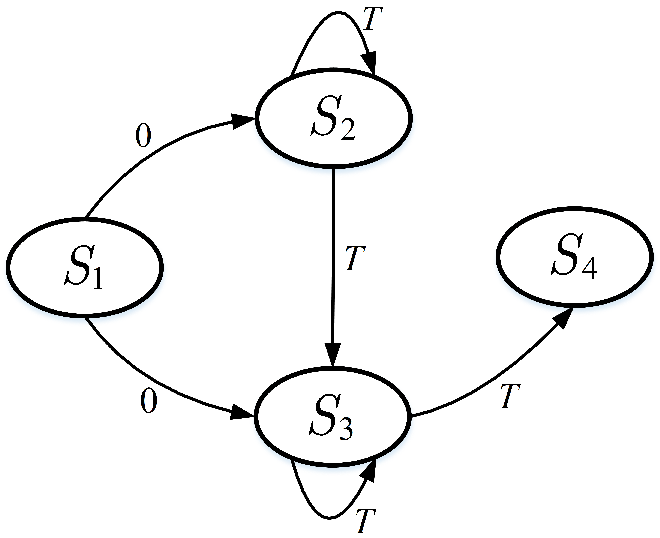}\\
  \caption{Illustration of the state transition process from the end instant of the ($j-1$)-th successful transmission to the end instant of the $j$-th successful transmission. The expressions along side the arrow lines denote the corresponding {durations} spent by the state {transitions}.}\label{Markov}
\end{figure}
$D_j$ denotes the time span between two neighboring updates, and it's worth noting that the state transition process between time {slots can} be described as a Markov chain with an absorbing wall as shown in Fig. \ref{Markov}, where state $S_1$ and state $S_4$ (i.e., absorbing state) denote
the end moment of the time slot of the ($j-1$)-th successful transmission and the $j$-th successful transmission, respectively, and $S_2$ and $S_3$ denote that $U_1$ has no packet to {transmit and has a packet} to be transmitted in buffer at the beginning of a slot, respectively. Thus, the state transfer probability matrix can be expressed as:
\begin{align}\label{}
\bm{\pi}&=
\begin{bmatrix}
 \bm{Q} & \bm{R} \\
\bm{0} & \bm{1}\\
\end{bmatrix},
\end{align}
where
\begin{align}\label{}
\bm{Q}\!=\!\begin{bmatrix}
0 & 1-\lambda & \lambda   \\
0 & 1-\lambda & \lambda\\
0 & 0 & 1-\tilde{P_1}
\end{bmatrix},
\bm{R}\!=\!\begin{bmatrix}
0  \\
0 \\
\tilde{P_1} \\
\end{bmatrix}.
\end{align}

The expression for $D_j$ which is the total time span from the start state $S_1$ to the absorbing state $S_4$ can be obtained as {follows}. First, from the Markov chain as shown in Fig. \ref{Markov}, it can be straightforwardly obtained that $D_j=(n-1)T$, where $n$ denotes the number of transition steps from state $S_1$ to state $S_4$. Hence, $\mathbb{E}\{D_j\}$ can be expressed by:
\begin{align}\label{NOMA-RT E(Dj)}
 \mathbb{E}\{D_j\}=& \mathbb{E}\{(n-1)T\},
\end{align}

Then, the following task is to evaluate $\mathbb{E}\{n\}$. Note that the expected number of steps from a transient state $S_i$ to the absorbing state can be obtained by a fundamental matrix $\bm{N}$ based on the absorbing Markov chain theory \cite{kemeny1976}, where the $(i,j)$-th entry of $\bm{N}$ denotes the expected number of visits containing a visit to the initial state before being absorbed from the transient state $S_i$ to the transient state $S_j$. Further, by summing the elements of the $i$-th row of the matrix $\bm{N}$, the expected number of steps from the transient state $S_i$ to the absorbing state can be obtained as follows:
\begin{align}\label{NOMA-RT vector v}
\mathbf{v}=\bm{N}\bm{c}=
\begin{bmatrix}
\frac{1}{\lambda}+\frac{1}{\tilde{P_1}}\\
\frac{1}{\lambda}+\frac{1}{\tilde{P_1}}\\
\frac{1}{\tilde{P_1}}
\end{bmatrix},
 \end{align}
where $\bm{N}=\sum_{k=0}^\infty \bm{Q}^k=(\bm{I}_3-\bm{Q})^{-1}$, and $\bm{I}_3$ is a $3 \times 3$ identity matrix and $\bm{c}$ is a column vector with all elements $1$.

Therefore, it can be straightforwardly obtained that $\mathbb{E}\{n\}=\bm{v}(1)$. Further, by taking $\mathbb{E}\{n\}$ into (\ref{NOMA-RT E(Dj)}), the expression for $\mathbb{E}\{D_j\}$ can be obtained as:
\begin{align}\label{TDMA-RT expectation of the interval of each update}
 \mathbb{E}\{D_j\}=T(\frac{1}{\lambda}+\frac{1}{\tilde{P_1}}-1).
\end{align}

Moreover, $\mathbb{E}\{D_j^2\}$ can be expressed as:
 \begin{align}\label{NOMA-RT E(Dj^2)}
 \mathbb{E}\{D_j^2\}=& \mathbb{E}\{[(n-1)T]^2\}\\\notag
=&T^2(\sigma^2(n)+ \mathbb{E}^2\{n\}-2 \mathbb{E}\{n\}+1),
\end{align}
where $\sigma^2(n)$ is the variance of $n$.

Note that based on {the} absorbing Markov chain theory \cite{kemeny1976}, the variance of the number of steps before being absorbed when starting from transient state $S_i$ can be obtained by the $i$-th entry of the vector $\bm{\varphi}$, which is given by:
\begin{align}\label{vector}
\bm{\varphi}=(2\bm{\bm{N}}-\bm{\bm{I}}_3)\bm{v}-
\begin{bmatrix}
(\frac{1}{\lambda}+\frac{1}{\tilde{P_1}})^2\\
(\frac{1}{\lambda}+\frac{1}{\tilde{P_1}})^2\\
(\frac{1}{\tilde{P_1}})^2
\end{bmatrix}
=
\begin{bmatrix}
s_1\\
s_2 \\
s_3
\end{bmatrix},
\end{align}
where
\begin{align}\label{}
s_1=&-(\lambda^2\tilde{P_1} - \lambda^2 + \lambda\tilde{P_1}^2 - P_1^2)/(\lambda^2\tilde{P_1}^2),\\
s_2=&-(\lambda^2\tilde{P_1} - \lambda^2 + \lambda \tilde{P_1}^2 - P_1^2)/(\lambda^2\tilde{P_1}^2),\\
s_3=&-(\tilde{P_1} - 1)/\tilde{P_1}^2.
\end{align}
Therefore, it can be straightforwardly obtained that $\sigma^2(n)=\bm{\varphi}(1)=s_1$. Then by taking $\mathbb{E}\{n\}$ and $\sigma^2(n)$ into (\ref{NOMA-RT E(Dj^2)}), the expression for $\mathbb{E}\{D_j^2\}$ can be obtained, which can be expressed as:
\begin{align}\label{NOMA-RT expectation of time squared per update interval}
&\mathbb{E}\{D_j^2\}\!=\!T^2(\frac{2(\lambda^2\!+\!\tilde{P_1}^2)\!-\!\lambda \tilde{P_1}(3\lambda\!+\!3\tilde{P_1}\!-\!\lambda\tilde{P_1}\!-\!2)}{\lambda^2\tilde{P_1}^2}).
\end{align}
Therefore, the proof is complete.
\end{IEEEproof}

{We} note that the expression for the average AoI achieved by the grant-free NOMA-RT scheme is much more complex than that of the NOMA-NRT scheme, making it difficult to find the optimal value of ${P}_{\text{TX}}$ for minimizing the average AoI. Thus, this paper relies on simulation results to  investigate the optimal ${P}_{\text{TX}}$ for the grant-free NOMA-RT scheme as will be demonstrated in the next section.

\begin{figure}[!t]
  	\centering
	\setlength{\abovecaptionskip}{0em}   
	\setlength{\belowcaptionskip}{-2em}
  \includegraphics[width=3in]{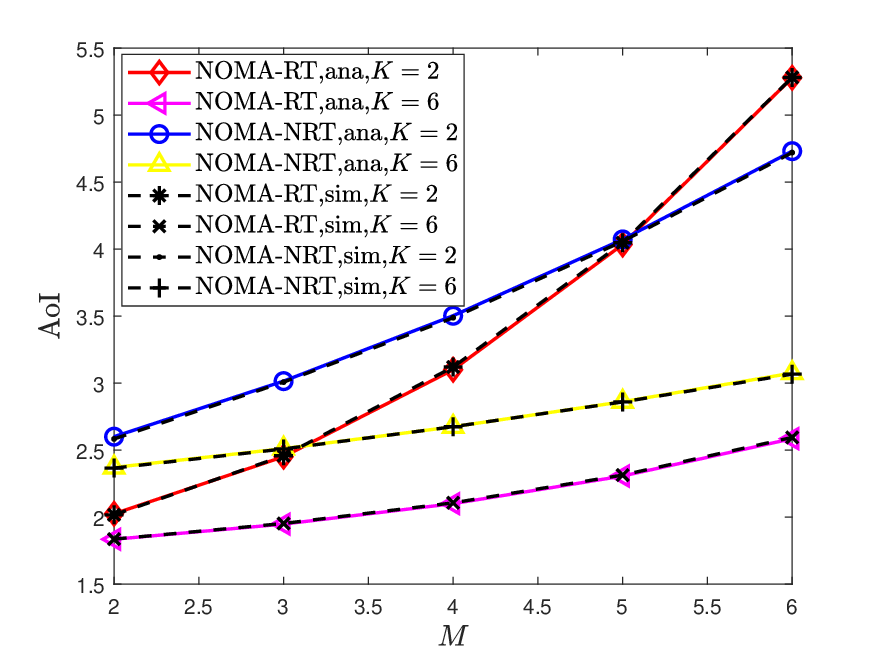}\\
  \caption{Average AoI achieved by the grant-free NOMA-NRT and NOMA-RT schemes. $\lambda=0.5$, $P_{\text{TX}}=0.5$, $P=20$ dB and $q_k=1/K$.}\label{anasim}
\end{figure}

\begin{figure}[!t]
  	\centering
	\setlength{\abovecaptionskip}{0em}  
	\setlength{\belowcaptionskip}{-2em}
  \includegraphics[width=3in]{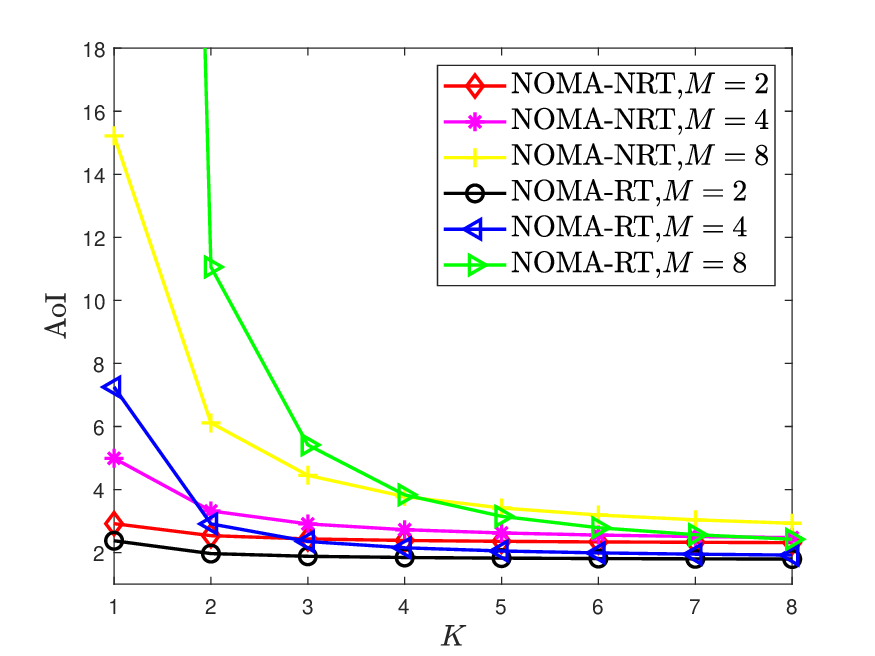}\\
  \caption{Impact of the number of received SNR levels on the average AoI achieved by the grant-free NOMA-NRT and NOMA-RT schemes. $\lambda=0.5$, $P_{\text{TX}}=0.5$, $P=20$ dB and $q_k=1/K$.}\label{receivedSNR}
\end{figure}

\section{Numerical Results}
In this section, numerical results are presented to verify the accuracy of the developed analysis, and also characterize the performance achieved by the considered grant-free NOMA-NRT and NOMA-RT schemes in terms of average AoI.

\begin{figure}[!t]
  \centering
  	\centering
  	\setlength{\abovecaptionskip}{0em}  
	\setlength{\belowcaptionskip}{-2em}
    \subfloat[NOMA-NRT]{\includegraphics[width=3in]{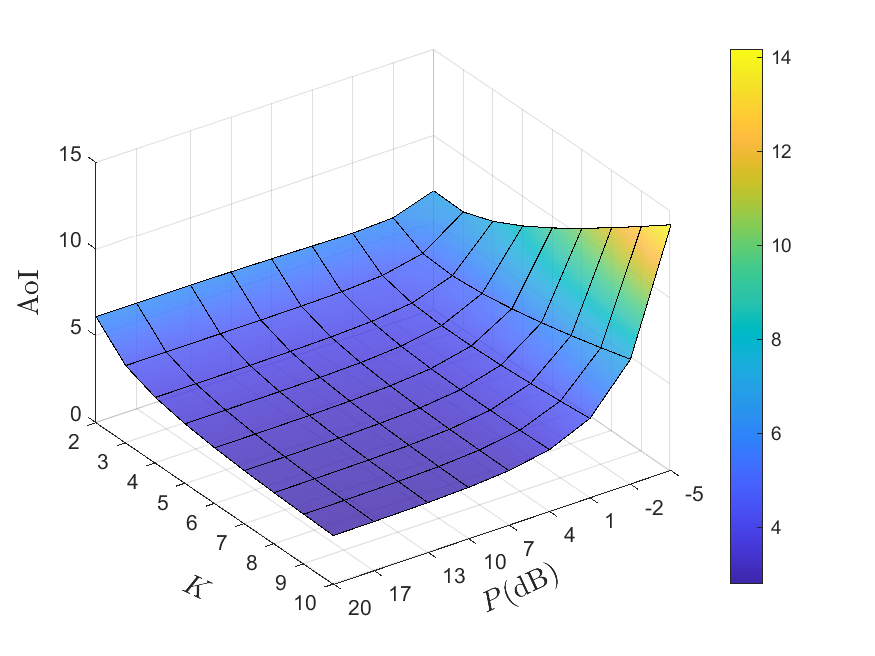}}\\
    \vspace{-1.0em}
  \subfloat[NOMA-RT]{ \includegraphics[width=3in]{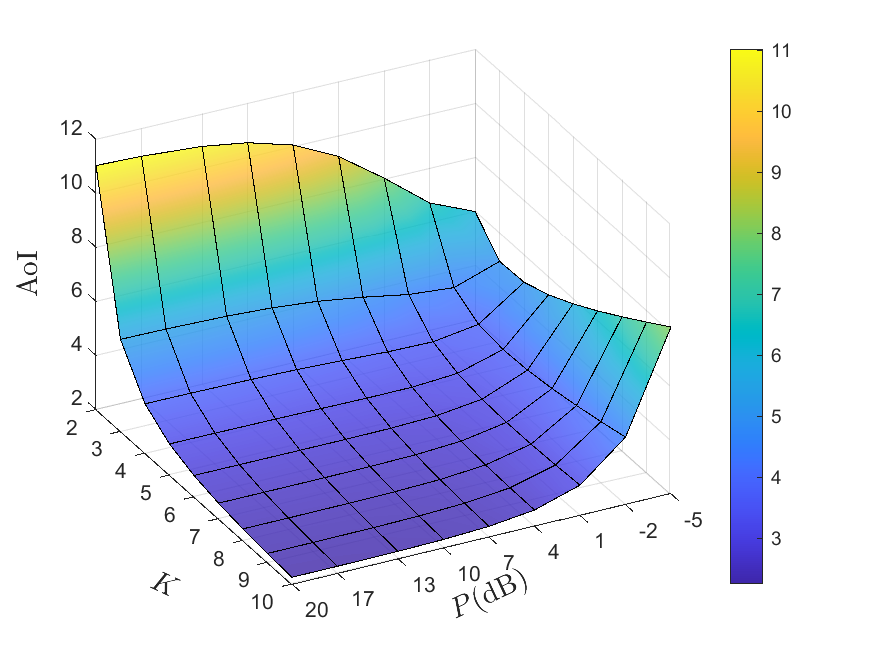}}
  \caption{Impact of the number of received SNR levels and transmission power on the average AoI achieved by the grant-free NOMA-NRT and NOMA-RT schemes. $M=8$, $\lambda=0.5$, $P_{\text{TX}}=0.5$, $R=0.2$ and $q_k=1/K$.}\label{receivedSNR_transmissionpower}
\end{figure}
\begin{table*}[!t]
    \centering
    \caption{Impact of the number of received SNR levels and transmission power on the average AoI achieved by the grant-free NOMA-NRT scheme. $M=8$, $\lambda=0.5$, $P_{\text{TX}}=0.5$, $R=0.2$ and $q_k=1/K$.}
    \renewcommand{\arraystretch}{1.3}
    \begin{tabular}{|c|c|c|c|c|c|c|c|c|c|}
        \hline
        \diagbox[width=8em, height=4em, dir=SE]{$K$}{AoI}{$P$(dB)} & -5 & -2 & 1 & 4 & 7 & 10 & 13 & 17 & 20 \\ \hline
        2     & 6.8236 & 6.0511 & 5.9473 & 5.9904 & 6.0415 & 6.0756 & 6.0949 & 6.1073 & 6.1116 \\ \hline
        3     & 6.7646 & 5.1606 & 4.6720 & 4.5221 & 4.4752 & 4.4596 & 4.4540 & 4.4513 & 4.4505 \\ \hline
        4     & 7.3874 & 5.0359 & 4.2508 & 3.9692 & 3.8603 & 3.8147 & 3.7943 & 3.7827 & 3.7790 \\ \hline
        5     & 8.2854 & 5.1797 & 4.1096 & 3.7097 & 3.5479 & 3.4774 & 3.4449 & 3.4262 & 3.4202 \\ \hline
        6     & 9.3341 & 5.4584 & 4.0977 & 3.5812 & 3.3688 & 3.2751 & 3.2316 & 3.2063 & 3.1982 \\ \hline
        7     & 10.4755 & 5.8203 & 4.1587 & 3.5233 & 3.2605 & 3.1440 & 3.0897 & 3.0582 & 3.0480 \\ \hline
        8     & 11.6756 & 6.2404 & 4.2670 & 3.5085 & 3.1943 & 3.0549 & 2.9899 & 2.9521 & 2.9399 \\ \hline
        9     & 12.9129 & 6.7041 & 4.4091 & 3.5227 & 3.1556 & 2.9929 & 2.9170 & 2.8729 & 2.8587 \\ \hline
        10    & 14.1730 & 7.2019 & 4.5773 & 3.5578 & 3.1359 & 2.9492 & 2.8623 & 2.8119 & 2.7957 \\ \hline
    \end{tabular}
    \label{NRT_K}
\end{table*}
\begin{table*}[!t]
    \centering
    \caption{Impact of the number of received SNR levels and transmission power on the average AoI achieved by the grant-free NOMA-RT scheme. $M=8$, $\lambda=0.5$, $P_{\text{TX}}=0.5$, $R=0.2$ and $q_k=1/K$.}
    \renewcommand{\arraystretch}{1.3}
    \begin{tabular}{|c|c|c|c|c|c|c|c|c|c|}
        \hline
        \diagbox[width=8em, height=4em, dir=SE]{$K$}{AoI}{$P$(dB)} & -5 & -2 & 1 & 4 & 7 & 10 & 13 & 17 & 20 \\ \hline
        2     & 5.9874 & 6.7062 & 8.0248 & 9.2250 & 10.0484 & 10.5345 & 10.7995 & 10.9672 & 11.0238 \\ \hline
        3     & 4.9508 & 4.3817 & 4.5249 & 4.8260 & 5.0756 & 5.2348 & 5.3248 & 5.3829 & 5.4026 \\ \hline
        4     & 4.9875 & 3.8310 & 3.5841 & 3.6121 & 3.6904 & 3.7524 & 3.7905 & 3.8160 & 3.8248 \\ \hline
        5     & 5.3280 & 3.7071 & 3.2246 & 3.1113 & 3.1057 & 3.1209 & 3.1340 & 3.1439 & 3.1476 \\ \hline
        6     & 5.8022 & 3.7490 & 3.0749 & 2.8629 & 2.8030 & 2.7889 & 2.7868 & 2.7872 & 2.7877 \\ \hline
        7     & 6.3478 & 3.8751 & 3.0232 & 2.7286 & 2.6268 & 2.5911 & 2.5778 & 2.5715 & 2.5697 \\ \hline
        8     & 6.9348 & 4.0520 & 3.0257 & 2.6549 & 2.5169 & 2.4631 & 2.4408 & 2.4290 & 2.4254 \\ \hline
        9     & 7.5465 & 4.2630 & 3.0623 & 2.6173 & 2.4456 & 2.3757 & 2.3454 & 2.3289 & 2.3237 \\ \hline
        10    & 8.1727 & 4.4984 & 3.1222 & 2.6032 & 2.3989 & 2.3137 & 2.2761 & 2.2552 & 2.2486 \\ \hline
    \end{tabular}
    \label{RT_K}
\end{table*}
Fig. \ref{anasim} shows the average AoI achieved by the grant-free NOMA-NRT and NOMA-RT schemes versus the number of sources $M$, respectively. The simulation results are obtained by averaging over $3\times10^5$ consecutive slots. It can be clearly observed from the figure that the simulation results perfectly match the analytical results, which verifies the accuracy of the developed analysis. Besides, it can be observed that as the number of users {grows}, the average AoIs achieved by both the grant-free NOMA-NRT and NOMA-RT schemes increase.

Fig. \ref{receivedSNR} shows the impact of the pre-configured number of received SNR levels on the average AoI achieved by the considered grant-free NOMA-NRT and NOMA-RT schemes, respectively. As shown in the figure, the average AoIs achieved by the grant-free NOMA-NRT and NOMA-RT schemes decrease as the number of received SNR levels $K$ increases. In addition, {an} increase of $K$ yields a sharper decrease in the average AoI for larger values of $M$. Moreover, the decreasing rate decays fairly fast as $K$ increases. 
In particular, when $K=1$, the NOMA schemes degrade to traditional OMA schemes in which only a single source can {transmit in} each time slot. It is obvious that the NOMA schemes significantly outperform their OMA counterparts in terms of average AoI, especially when the number of users is relatively large. Besides, it is shown that the retransmission schemes achieve lower average AoIs only when $K$ is sufficiently large.

Fig. \ref{receivedSNR_transmissionpower} shows the impact of the transmission power and the number of received SNR levels on the average AoI achieved by the grant-free NOMA-NRT and NOMA-RT schemes. From Fig. \ref{receivedSNR_transmissionpower} (a) and Table \ref{NRT_K}, it can be observed that at high values of $P$, the AoI achieved by the grant-free NOMA-NRT scheme decreases with $K$. 
However, at low values of $P$, the AoI first decreases and then increases as $K$ increases. 
The reasons can be explained as follows. 
{The AoI first decreases with $K$} because of a reduction in the probability of collisions as $K$ increases.
However, when $K$ becomes larger, due to the limited power budget, the actual probability that a source transmits will decrease, which yields a higher AoI. 
In addition, from Fig. \ref{receivedSNR_transmissionpower} (a) and Table \ref{NRT_K}, it can be observed that when $K > 2$, increasing $P$ gradually reduces the AoI achieved by the grant-free NOMA-NRT scheme. Differently, when $K = 2$, the AoI first decreases and then increases with $P$. This is because increasing $P$ improves the probability of successful transmission, which is helpful in reducing AoI on the one hand, and meanwhile increases the probability of collisions, which counteracts the reduction {of} AoI on the other hand. 
As shown in Fig. \ref{receivedSNR_transmissionpower} (b), a very different observation for NOMA-RT scheme compared to NOMA-NRT scheme is that
when $K$ is low while $P$ is high, the average AoIs achieved by NOMA-RT are much higher than those achieved by NOMA-NRT. The reason is because 
NOMA-RT is more susceptible to collisions  when $K$ is low while $P$ is high.  

\begin{figure}[!t]
  \centering
  	\centering
  	\setlength{\abovecaptionskip}{0em}  
	\setlength{\belowcaptionskip}{-2em}
    \subfloat[NOMA-NRT]{\includegraphics[width=3in]{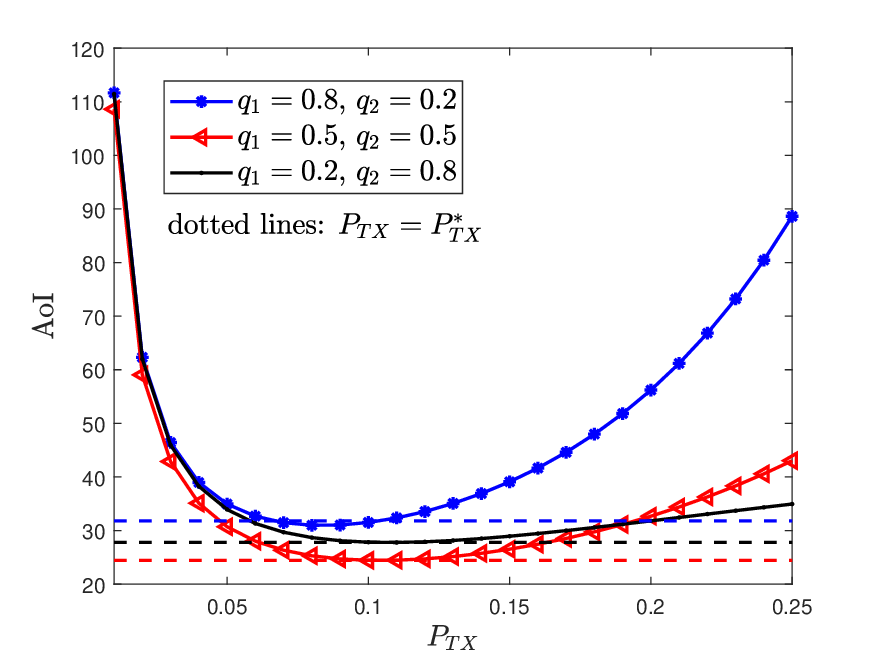}}\\
    \vspace{-1.2em}
  \subfloat[NOMA-RT]{ \includegraphics[width=3in]{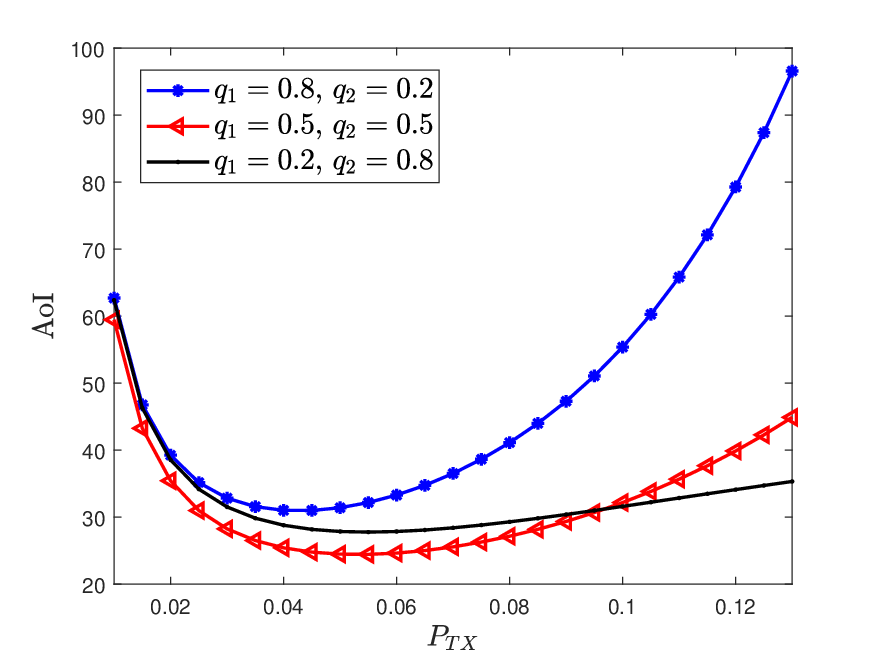}}
  \caption{Impact of the attempted transmission probability on the average AoI achieved by the grant-free NOMA-NRT and NOMA-RT schemes. $M=32$, $K=2$, $P=20$ dB and $\lambda=0.5$.}\label{attempted_AoI}
\end{figure}
Fig. \ref{attempted_AoI} shows the impact of the probability of attempted transmission on the average AoI achieved by the grant-free NOMA-NRT and NOMA-RT schemes. As shown in the figure, the AoIs achieved by both NOMA-RT and NOMA-NRT schemes first decrease and then increase with $P_{\text{TX}}$. Moreover, it can be observed that the optimal values of $P_{\text{TX}}$ for the NOMA-NRT scheme perfectly match the dashed lines which are based on the analytical results as shown in Corollary $1$. Therefore, the accuracy of Corollary $1$ {is} verified. 
In addition, it is shown that the choice of $q_k$ significantly affects the optimal $P_{\text{TX}}$. 
For example, as shown in the figure, when $K=2$, the optimal $P_{\text{TX}}$ increases with $q_1$. 
Furthermore, the results in the figure demonstrate that the minimum average AoI values decrease as the discrepancy between $q_1$ and $q_2$ narrows.
The underlying reason is that a greater disparity between $q_1$ and $q_2$ leads to either an elevated collision probability or a reduced successful transmission probability. 

\begin{figure}[!t]
  \centering
  	\centering
  	\setlength{\abovecaptionskip}{0em}  
	\setlength{\belowcaptionskip}{-1em}
    \subfloat[NOMA-NRT]{\includegraphics[width=3in]{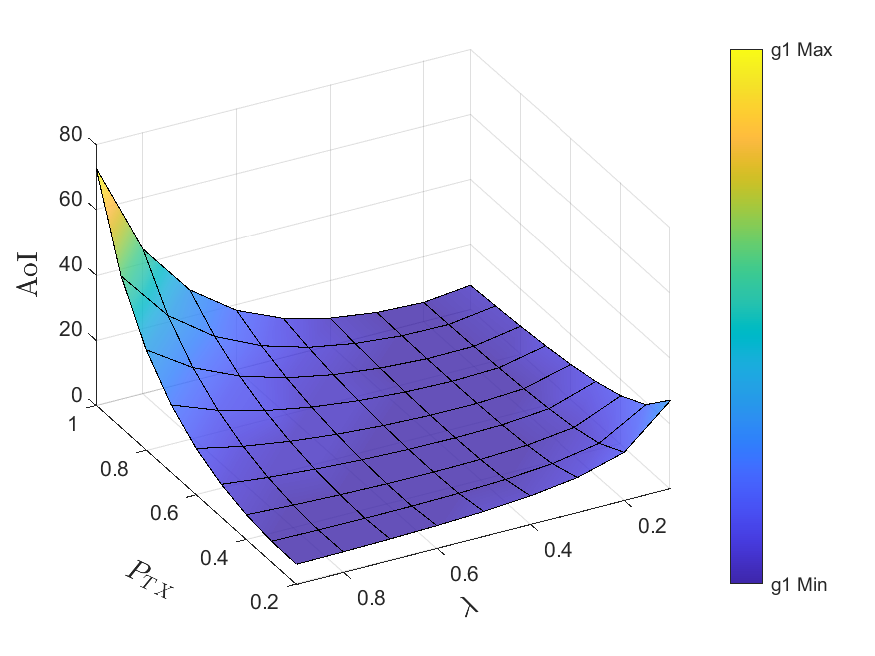}}\\
    \vspace{-1.2em}
  \subfloat[NOMA-RT]{ \includegraphics[width=3in]{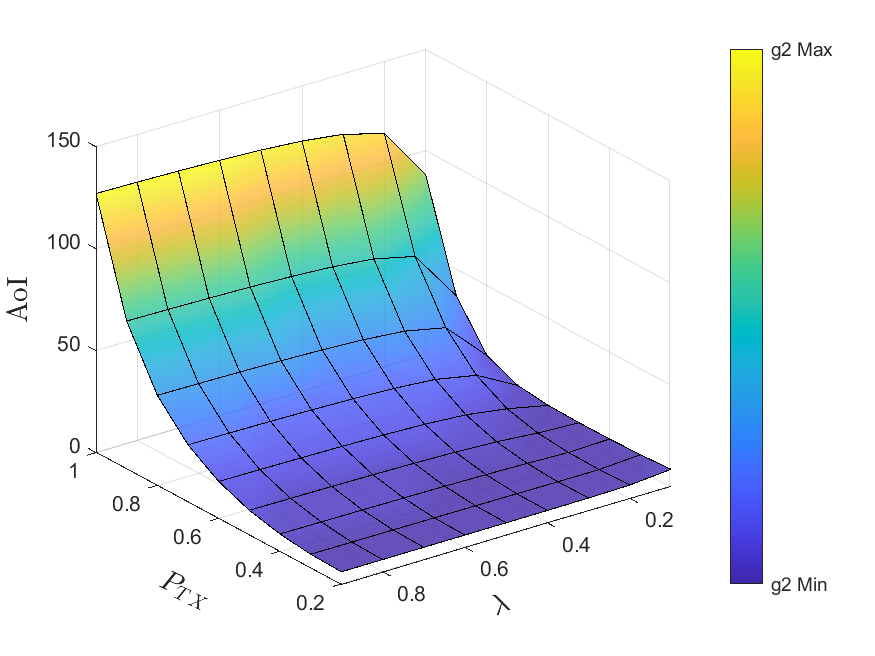}}
  \caption{Impact of data arrival rates and attempted transmission probabilities on the average AoI achieved by the grant-free NOMA-NRT and NOMA-RT schemes. $M=8$, $P=20$ dB, $K=2$ and $q_k=1/K$.}\label{rates_attempted_K2}
\end{figure}
\begin{figure}[!t]
  \centering
  	\centering
 	\setlength{\abovecaptionskip}{0em}  
	\setlength{\belowcaptionskip}{-1.5em}
        \vspace{-1.2em}
    \subfloat[NOMA-NRT]{\includegraphics[width=3in]{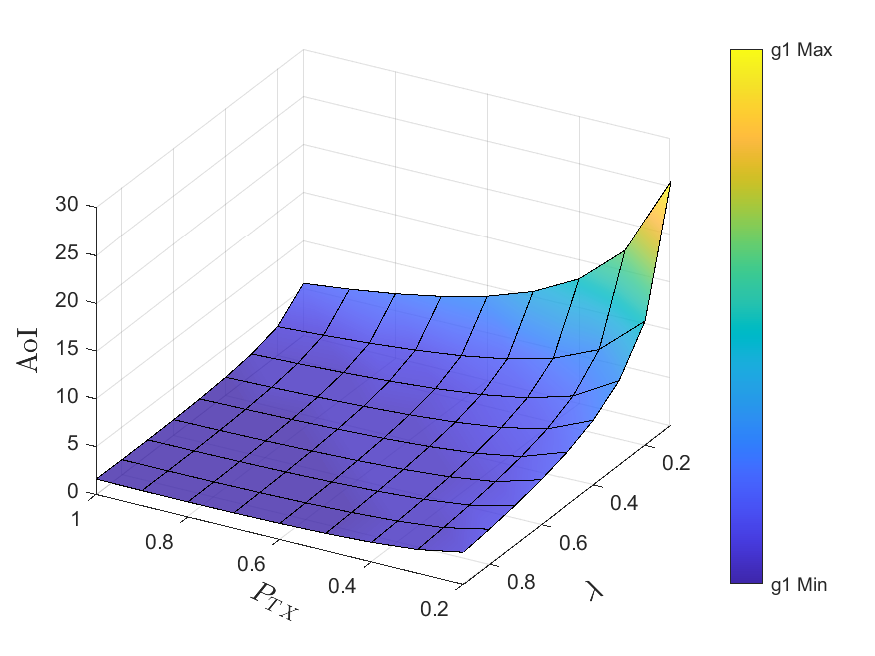}}\\
  \subfloat[NOMA-RT]{ \includegraphics[width=3in]{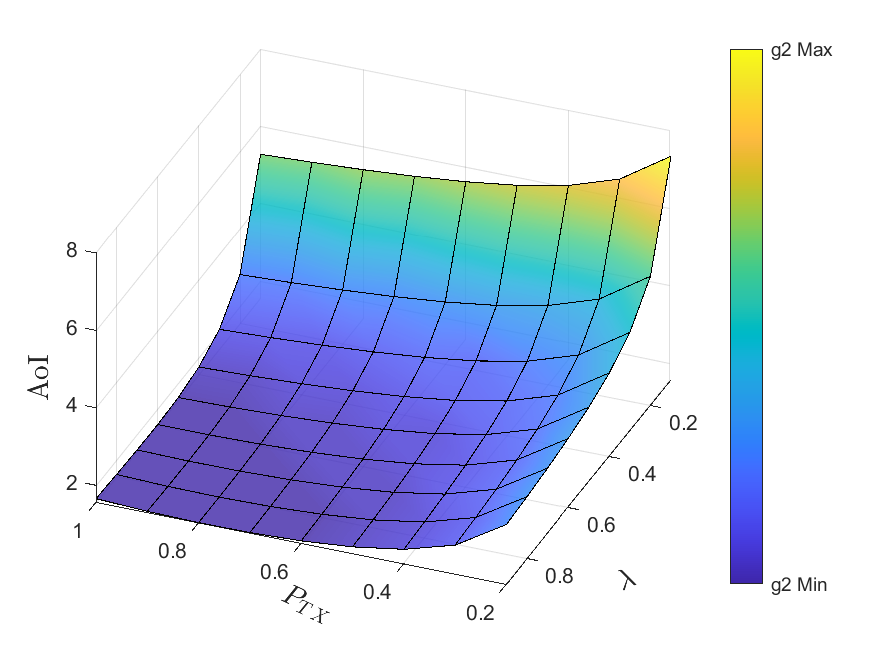}}
 \caption{Impact of data arrival rates and attempted transmission probabilities on the average AoI achieved by the grant-free NOMA-NRT and NOMA-RT schemes. $M=8$, $P=20$ dB, $K=16$ and $q_k=1/K$.}\label{rates_attempted_K16}
\end{figure}
Fig. \ref{rates_attempted_K2} and Fig. \ref{rates_attempted_K16} show the impact of data arrival rate and attempted transmission probability on the average AoI achieved by the grant-free NOMA-NRT and NOMA-RT schemes with $K=2$ and $K=16$, respectively.
As shown in Fig. \ref{rates_attempted_K2} (a), when $K=2$, both low $\lambda$ with low
$P_{\text{TX}}$ and high $\lambda$ with high $P_{\text{TX}}$ lead to a higher average AoI achieved by the grant-free NOMA-NRT scheme. 
The reasons can be explained as follows. When both $\lambda$ and $P_{\text{TX}}$ are low, the reduction in the average AoI is limited by insufficient transmission opportunities, which can be significantly improved by increasing $\lambda$ and (or) $P_{\text{TX}}$. In contrast, 
when both $\lambda$ and $P_{\text{TX}}$ reach sufficiently high levels, the reduction in the average AoI is mainly constrained by elevated collision probabilities, which will be exacerbated by the increase of $\lambda$ and $P_{\text{TX}}$. In contrast, for the grant-free NOMA-RT scheme, as shown in Fig. \ref{rates_attempted_K2} (b), when $P_{\text{TX}}$ is fixed, increasing $\lambda$ further has {insignificant} impact on the average AoI, which is due to the incorporated retransmission mechanism. 
Interestingly and differently, when $K$ becomes larger, as illustrated in Fig. \ref{rates_attempted_K16}, the increase of $\lambda$ and $P_{\text{TX}}$ are both beneficial for the decrease of the average AoI for both the NOMA-RT and NOMA-NRT schemes. Because when 
$K$ is sufficiently large, the collision probability decreases significantly, making the transmission opportunity the dominant factor to reduce AoI. 

\begin{figure}[!t]
  \centering
  	\centering
  	\setlength{\abovecaptionskip}{0em}  
	\setlength{\belowcaptionskip}{-2em}
    \subfloat{\includegraphics[width=3in]{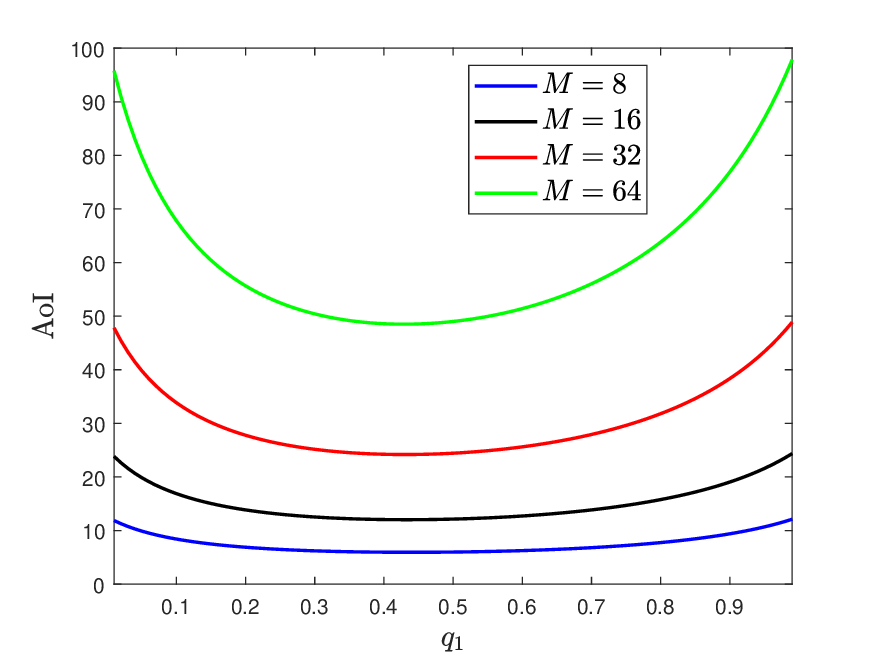}}
  \caption{Impact of $q_k$ on the average AoI achieved by the grant-free NOMA-NRT scheme. $K=2$, $P=20$ dB, $P_{\text{TX}}=P_{\text{TX}}^*$ and $\lambda=0.4$.}\label{qk_AoI}
\end{figure}
\begin{figure}[!t]
  \centering
  	\centering
  	\setlength{\abovecaptionskip}{0em}  
	\setlength{\belowcaptionskip}{-1em}
    \subfloat{\includegraphics[width=3in]{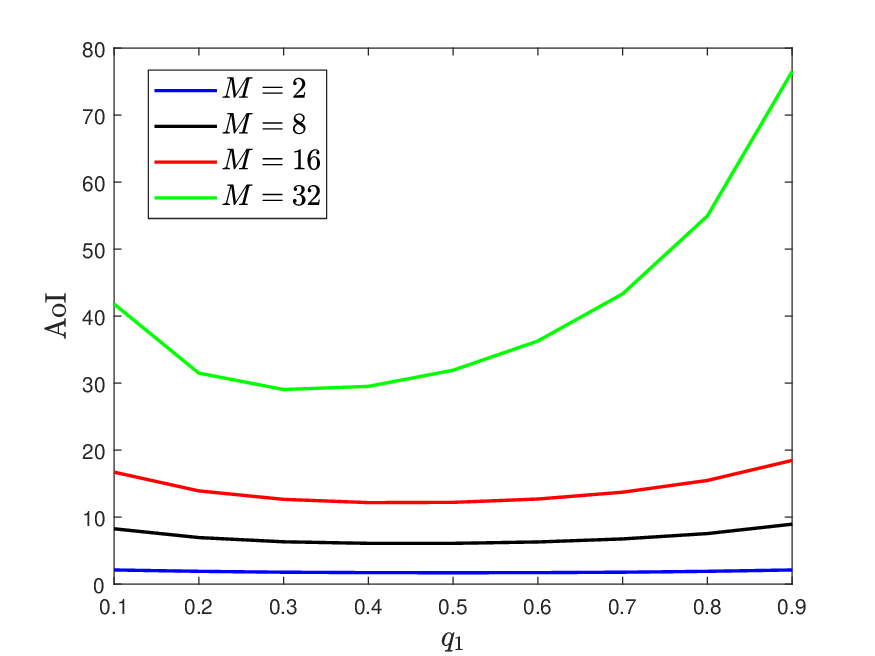}}
  \caption{Impact of $q_k$ on the average AoI achieved by the grant-free NOMA-RT scheme, where $P_{\text{TX}}$ is optimized via exhaustive searching. $K=2$, $P=20$ dB and $\lambda=0.4$.}\label{qk_AoI_RT}
\end{figure}
\begin{figure}[!t]
  \centering
  	\centering
  	\setlength{\abovecaptionskip}{0em}  
	\setlength{\belowcaptionskip}{-2em}
    \subfloat[M=2]{\includegraphics[width=3in]{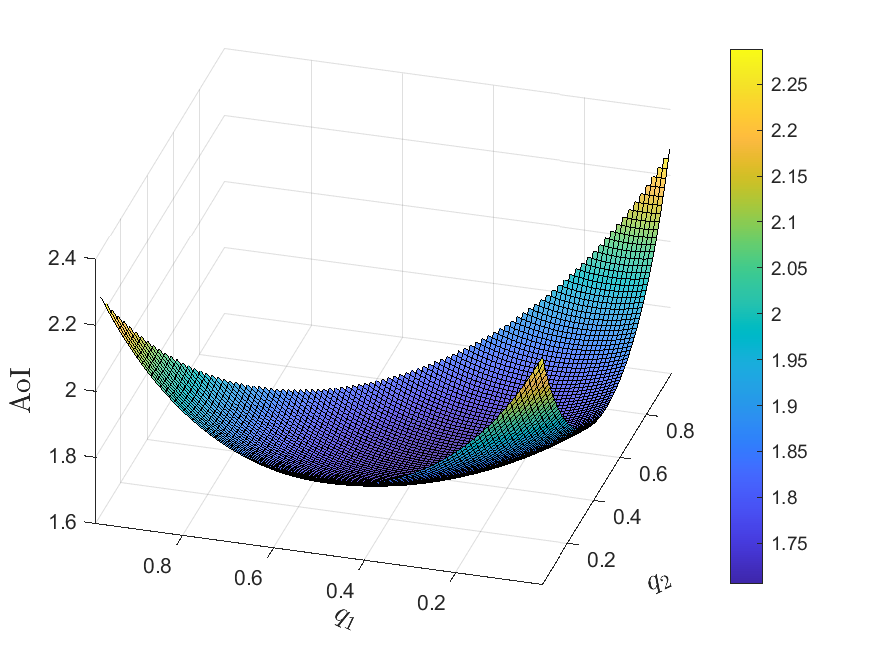}}\\
  \subfloat[M=8]{ \includegraphics[width=3in]{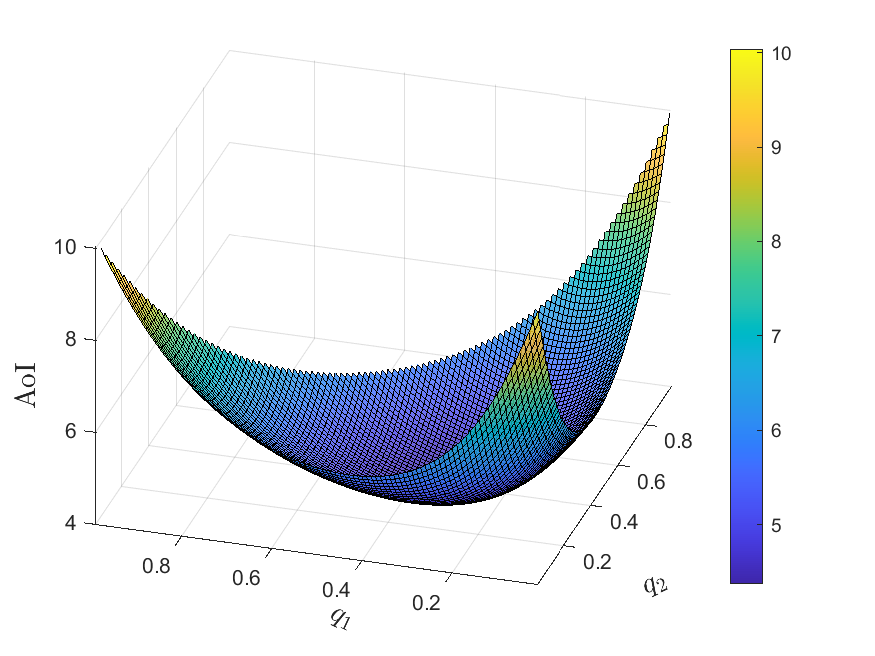}}
  \caption{Impact of $q_k$ on the average AoI achieved by the grant-free NOMA-NRT scheme, where $P_{\text{TX}}$ is optimized via exhaustive searching. $K=3$, $P=20$ dB and $\lambda=0.4$.}\label{qk_AoI_K3}
\end{figure}
\begin{figure}[!t]
  \centering
  	\centering
  	\setlength{\abovecaptionskip}{0em}  
	\setlength{\belowcaptionskip}{-2em}
    \subfloat[M=2]{\includegraphics[width=3in]{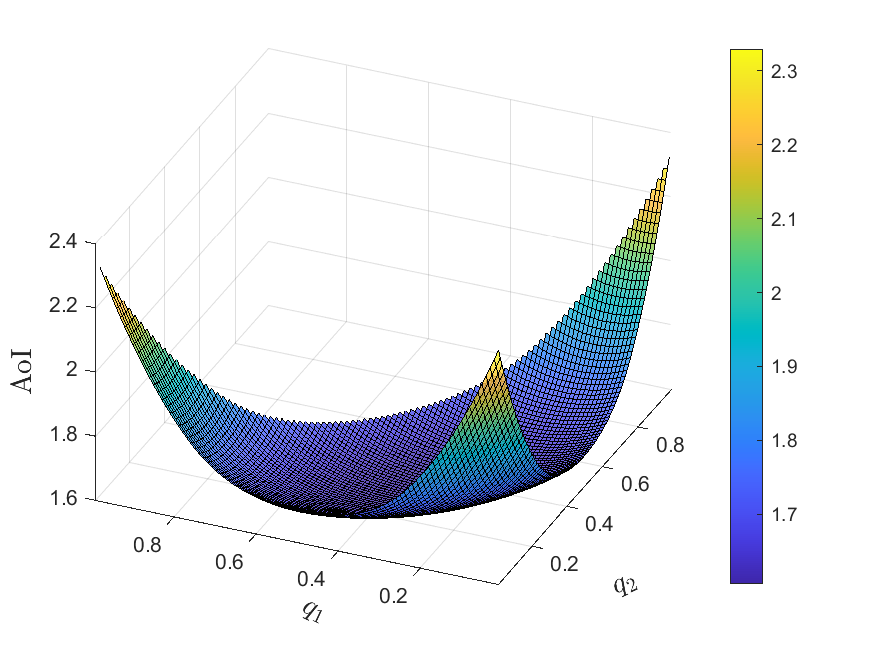}}\\
    \vspace{-1em}
  \subfloat[M=8]{ \includegraphics[width=3in]{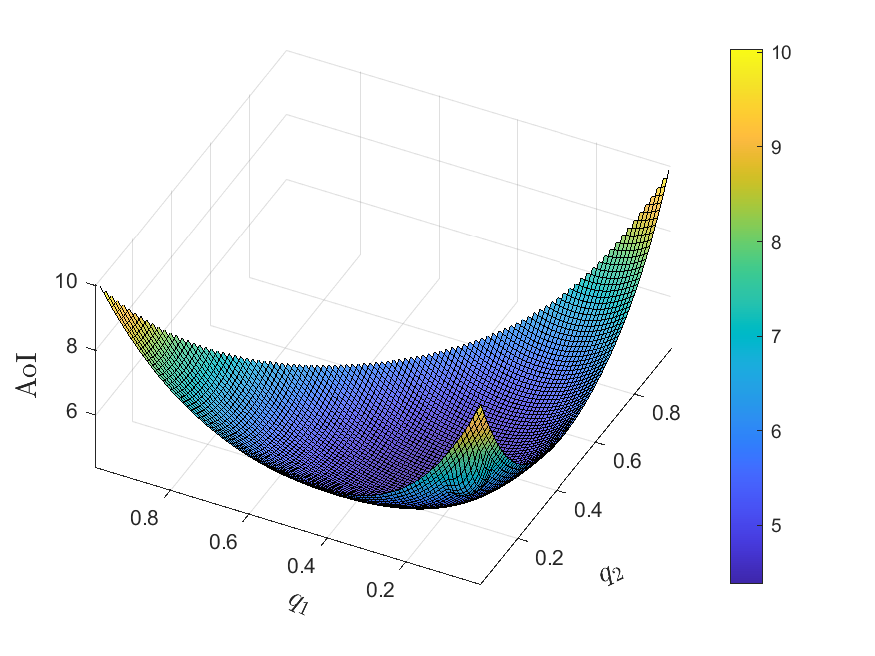}}
  \caption{Impact of $q_k$ on the average AoI achieved by the grant-free NOMA-RT scheme, where $P_{\text{TX}}$ is optimized via exhaustive searching. $K=3$, $P=20$ dB and $\lambda=0.4$.}\label{qk_AoI_K3_RT}
\end{figure}
Fig. \ref{qk_AoI} and  Fig. \ref{qk_AoI_RT} show the impact of $q_k$ on the average AoI achieved 
by the grant-free NOMA-NRT and NOMA-RT schemes with $K=2$. As can be seen in the figures, the average AoIs initially decrease, then keep almost invariant, and finally increase, as $q_1$ increases. Besides, the interval of $q_1$ where the average AoI keeps almost invariant becomes larger as the total number of sources increases.  
{These presented numerical results demonstrate that} the optimal values of $q_1$ in NOMA-RT and NOMA-NRT schemes are  different. In particular, as shown in  Fig. \ref{qk_AoI} and  Fig. \ref{qk_AoI_RT}, the optimal values of 
$q_1$ for the grant-free NOMA-NRT scheme and NOMA-RT schemes are different. Specifically, the optimal $q_1$ for
the grant-free NOMA-NRT scheme is about $0.43$ while that for the grant-free NOMA-RT scheme is about $0.3$. 
Fig. \ref{qk_AoI} and  Fig. \ref{qk_AoI_RT} also show that the grant-free NOMA-RT scheme achieve lower average AoI compared to the corresponding grant-free NOMA-RT scheme, especially when $M$ is large. Furthermore, 
Figs. \ref{qk_AoI_K3} and \ref{qk_AoI_K3_RT} show the impact of $q_k$ on the average AoI achieved by the grant-free NOMA-NRT and NOMA-RT schemes with $K=3$. From the two figures, it can be {observed} that when $K=3$, the minimum AoI achieved by the schemes with and without retransmission are almost the same. However, numerical results show that for the same pairs of $(q_1,q_2,q_3)$, the optimal $P_{\text{TX}}$ of the NOMA-NRT and NOMA-RT schemes are very different.
\vspace{-1.5em}
\section{Conclusions}
In this paper, the application of NOMA to grant-free transmissions for reducing AoI in uplink status update systems has been investigated, by taking into account the randomness of status updating arrivals. Transmission strategies with and without retransmission have been considered. 
A rigorous analytical framework has been developed to evaluate the average AoI achieved by the considered grant-free NOMA-RT and NOMA-NRT schemes, respectively. Extensive simulation results have been provided to verify the developed analysis, and also demonstrate the superior performance of 
NOMA assisted grant-free schemes. It has been shown that by applying NOMA the average AoI can be significantly reduced compared to OMA based schemes. 
Moreover, it has also been shown that the retransmission schemes cannot always outperform the schemes without retransmissions in the considered scenario. 
\vspace{-0.8em}
\bibliographystyle{IEEEtran}
\bibliography{IEEEabrv,ref}
\end{document}